\documentclass[amsmath,amssymb,twocolumn]{revtex4}

\usepackage{dcolumn}
\usepackage{bm}
\usepackage{amsmath,mathrsfs}
\usepackage{graphicx,epsfig}
\usepackage{hyperref}
\usepackage{epsf}
\usepackage{color}
\usepackage{subcaption}
\begin{document}


\title{{\bf Wormholes in  $f(T,\mathcal{T})$ gravity}}

\author{ F. Parsaei }\email{fparsaei@gmail.com}
\author{S. Rastgoo}\email{rastgoo@sirjantech.ac.ir}

\affiliation{ Physics Department , Sirjan University of Technology, Sirjan 78137, Iran.}

\date{\today}


\begin{abstract}
 \par This study aims to investigate the physical properties of wormhole geometry within the context of $f(T,\mathcal{T})$ gravity, which serves as a teleparallel formulation of general relativity. We study a linear model,  $f(T,\mathcal{T})=\alpha T+\beta\mathcal{T}$, to explore  traversable wormholes. A linear equation of state is utilized for radial  pressure, leading to a power-law shaped function. It was found that the violation of energy conditions, depends on the $\alpha$ and $\beta$ parameters. A diverse array of intriguing wormhole solutions was identified, contingent upon the specific model parameters employed. It is demonstrated that isotropic wormhole solutions cannot be attained within this framework. Additionally, solutions characterized by a variable equation of state parameter are introduced. A comparative analysis of wormhole solutions in the context of $f(T,\mathcal{T})$ and curvature-based gravity, specifically $f(R,T)$, is also provided. \\
\end{abstract}

\maketitle
\section{Introduction}
The presence of wormhole geometries represents one of the most compelling topics in modern cosmology. Wormholes serve as passage-like topological constructs that link two distant points within the same universe or across distinct universes via a shortcut known as a tunnel or bridge \cite{WH}.  The idea of wormholes was first introduced by Flamm, who developed the isometric embedding of the Schwarzschild solution \cite{flamm}. The traversable wormhole solution was initially introduced by Morris and Thorne. This concept significantly differed from the earlier hypotheses proposed by Einstein and Rosen in 1935, which resulted in what is now recognized as the Einstein-Rosen bridge \cite{Rosen}. Wormholes may be of a size adequate for humanoid travelers and could potentially enable time travel \cite{WH}. Morris and Thorne indicated that the theoretical fluid contained within the wormhole exhibits negative energy, thus violating the null energy condition (NEC) \cite{Visser}. The breach of the NEC is regarded as a fundamental prerequisite for the existence of wormhole geometries. So, the presence of a certain exotic fluid, which is a speculative form of matter, is essential to disrupt the NEC in general relativity (GR).

One of the most captivating discoveries in modern cosmology is the accelerated expansion of the Universe. This phenomenon has attracted considerable interest from scholars in the discipline. To describe this phenomenon, scientists have proposed a range of ideas that can be categorized into two distinct types of modified propositions: those concerning modified matter and those related to modified curvature. Phantom fluid can be considered in the first category. In the wake of this finding, comprehensive studies on phantom wormholes that violate energy conditions (ECs) have been undertaken in the literature \cite{phantom, phantom2, phantom1}. On the other hand, the importance of the Casimir effect in the context of traversable wormholes pertains to the presence of exotic matter, which contravenes the NEC \cite{cas}.
The primary challenge related to wormhole theory within the context of GR is the minimization of exotic matter requirements. In this theoretical framework, exotic matter is restricted to a specific region of spacetime. Notable instances of this include thin shell wormholes \cite{cut, cut1, cut2, cut3}, those with a variable equation of state (EoS)\cite{Remo, variable}, and wormholes described by a polynomial EoS \cite{foad}.

The second previously noted consideration, for examining the accelerated expansion of the Universe, involves proposing a modification to the action of Einstein's GR by incorporating additional degrees of freedom. This leads to the presentation of modified gravities in the literature. Modified gravity is another alternative to resolve the problem of exotic matter in the wormhole theory.  Numerous researchers have explored the physical characteristics of wormholes across different scenarios. Wormholes are studied in Braneworld \cite{b, b1, b2, b3}, Born-Infeld theory \cite{Bo, Bo1}, quadratic gravity \cite{quad, quad1}, Einstein-Cartan gravity \cite{Cartan, Cartan1, Cartan2}, Rastall–Rainbow gravity \cite{RaR, RaR1},  $f(Q)$ gravity \cite{fq, fq2, fq3, fq4, fq44, fq5, fq6}, $f(R)$ gravity \cite{Nojiri, fR0, fR11, fR22, fR33, fR44,fR55}, Ricci inverse gravity \cite{inverse}, and $f(R,T)$ gravity \cite{Azizi, Moa, Zub, Shar, Cha, Sharif, Rosa, Moa2,   fr2, fr3, Yousaf, Sha,  god, Bha, Ban, Sarkar, Chau, SR1}. Some of these modified theories can resolve the problem of exotic matter for  wormholes. In modified gravitational theories, it is common to extend the Einstein-Hilbert action of GR, beginning with the curvature-based description of gravity. Nevertheless, an intriguing alternative class of modified gravity emerges when one considers the action derived from the equivalent formulation of GR that incorporates torsion. Einstein  developed the "Teleparallel Equivalent of General Relativity" (TEGR), in which the gravitational field is characterized by the torsion tensor rather than the curvature tensor. The $f(T)$ gravity, referred to as the teleparallel theory of gravity, was initially employed by Einstein in his efforts to formulate a unified theory encompassing both gravity and electromagnetism. There exist intriguing articles regarding wormholes that utilize this formalism \cite{FT, FT1, FT2}.

The generalization of GR to $f(R)$ gravity has been paralleled by the extension of teleparallel gravity to $f(T)$ gravity. In a manner analogous to the generalization of $f(R,T)$ gravity, where $T$ represents the trace of the stress-energy tensor, $f(T)$ gravity can be extended to $f(T,\mathcal{T})$ gravity, with $\mathcal{T}$  denoting the trace of the stress-energy tensor and $T$ represent the torsion \cite{Harko}. The $f(T,\mathcal{T})$ gravity framework employs a TEGR, wherein the Weitzenböck connection is utilized in place of the torsionless Levi-Civita connection \cite{Harko}. In this context, the fundamental dynamical entities are represented by four linearly independent vierbein elements. Notably, the Weitzenböck connection is devoid of curvature and serves to characterize the torsion present within a manifold.

The literature has explored the cosmological phenomena within the context of  $f(T,\mathcal{T})$ gravity \cite{t1, t2, t3, t4, t5, t6, t7, t8, t9, t10, t11, t12, t13, t14, t15, t16}. Stars are investigated in this scenario \cite{star11, star13, star14}. Traversable Wormholes in the extended teleparallel theory
of gravity with matter coupling have been investigated by Mustafa et al. \cite{Must1}. They examined potential scenarios for wormhole geometries by employing Gaussian and Lorentzian noncommutative distributions to derive the precise shape function associated with these geometries \cite{Must1}. In \cite{Ditta}, the shape function is determined within the context of embedded class-1 spacetime in $f(T,\mathcal{T})$ gravity. The authors have utilized both diagonal and off-diagonal tetrads to establish a comparison by examining the validity region of ECs in embedded class-1 spacetime \cite{Ditta}. Some exact wormhole solutions in the context of  $f(T,\mathcal{T})$ are presented in \cite{Err}, which violate the ECs in the vicinity of the wormhole throat. Rizwan et al. have explored the properties of Casimir wormholes, incorporating modifications derived from the generalized uncertainty principle (GUP) within the context of matter-coupled teleparallel gravity \cite{Riz}. In another work, Chalavadi et al. have studied the characteristics of Casimir wormholes through the lens of  $f(T,\mathcal{T})$ gravity, while also integrating the GUP into their analysis \cite{Chal}. Their findings indicate that the resultant wormhole solutions display anisotropic properties and contravene the NEC, suggesting the existence of exotic matter. In \cite{Must2}, an investigation is conducted into the potential existence of generalized wormhole models within the galactic halo region through the analysis of particular dark matter models. The existing literature indicates that all proposed wormhole solutions, in the context of $f(T,\mathcal{T})$  gravity, contravene the classical ECs. Therefore, the primary objective of this paper is to identify wormhole solutions that comply with these ECs. In pursuit of this objective, we reformulate the field equations within the context of $f(T,\mathcal{T})$ gravity to establish a clear connection between the field equations in the GR framework and those in the $f(T,\mathcal{T})$ setting. Subsequently, we employ a linear and variable EoS to explore solutions that satisfy the ECs.

The organization of our paper is as follows: In Sec. \ref{sec2}, we investigate the criteria and equations that characterize wormholes. Subsequently, we provide a concise overview of $f(T,\mathcal{T})$ theory in conjunction with the classical ECs. In Sec.\ref{sec3}, we apply the field equations to derive the shape function within the context of $f(T,\mathcal{T})$ gravity, offering solutions that comply with the ECs. This section also encompasses an analysis of the physical properties related to these solutions. We investigate solutions with variable EoS  in Sec.\ref{sec4}. Finally, we present our final remarks in the concluding section. Throughout this manuscript, we have adopted the framework of gravitational units, specifically setting $c = 8 \pi G = 1$.

\section{Basic formulation of wormhole and $f(T,\mathcal{T})$ gravity} \label{sec2}
The line element of a metric that is both static and spherically symmetric can be expressed as follows
\begin{equation}\label{1}
ds^2=-U(r)dt^2+\frac{dr^2}{1-\frac{b(r)}{r}}+r^2(d\theta^2+\sin^2\theta,
d\phi^2)
\end{equation}
where $U(r)=\exp (2\phi(r))$.
The metric function $b(r)$ is known as the shape function, which characterizes the geometry of the wormhole. In this context, $\phi(r)$ is known as the redshift function, which can be utilized to ascertain the redshift of the signal as perceived by a distant observer.
The throat condition states that at the throat
\begin{equation}\label{2}
b(r_0)=r_0
\end{equation}
where  $r_0$ is the wormhole throat. Furthermore, two more conditions must be met to guarantee the existence of a traversable wormhole,
\begin{equation}\label{3}
b'(r_0)<1
\end{equation}
and
\begin{equation}\label{4}
b(r)<r,\ \ {\rm for} \ \ r>r_0.
\end{equation}
Equation (\ref{3})  is well-known as the flaring-out condition. This equation provides the violation of the NEC in the background of GR. Finally, the shape function must satisfy the asymptotically flat condition,
\begin{equation}\label{5}
\lim_{r\rightarrow \infty}\frac{b(r)}{r}=0,\qquad   \lim_{r\rightarrow \infty}U(r)=1,
\end{equation}
for a standard Morris-Thorne wormhole. To enhance simplicity, we focus on solutions that incorporate a constant redshift function. The vanishing redshift function represents a fundamental and frequently employed scenario in the study of wormholes within the framework of GR or its modified theories of gravity. The vanishing redshift function indicates the lack of tidal forces.
 This article investigates an anisotropic fluid represented by the tensor $\mathcal{T}^{\mu}_{\nu}=diag[-\rho, p, p_t, p_t]$, where $\rho$ signifies the energy density, $p$ represents the radial pressure, and $p_t$ indicates the tangential pressure, respectively.

We have a specific interest in $f(T,\mathcal{T})$ gravity, in which the Lagrangian is defined as an arbitrary function of the  torsion scalar
$T$ and the trace of the energy-momentum tensor $\mathcal{T}$. We will define
\begin{equation}\label{6a}
{e^i}_{ \mu}{e_i}^{ \mu}=\delta^\mu_\mu,
\end{equation}
where the metric is specified by
 \begin{equation}\label{6b}
g_{\mu \nu}=\eta_{ij}{e^i}_{\mu}{e^j}_{\nu},
\end{equation}
In this context, $\eta_{ij}= diag(-1, 1, 1, 1)$  represents the standard Minkowski metric, which serves geometrically as the metric of the tangent space.
The Einstein-Hilbert action integral for the $f(T,\mathcal{T})$ modified
gravity is
\begin{equation}\label{6}
S=\int\frac{1}{2}f(T,\mathcal{T})\sqrt{-g}\;d^{4}x+\int L_{m}\sqrt{-g}\;d^{4}x.
\end{equation}
where $f(T,\mathcal{T})$ is a general function  of $T$  and $\mathcal{T}$, $\sqrt{-g}$ is the determinant
of the metric, and $L_m$ represents the matter Lagrangian density \cite{Harko}. This geometrically altered action incorporates a well-defined function of $f(T,\mathcal{T})$, which substitutes the traditional Ricci scalar ($R$) in the Einstein-Hilbert action. The fundamental components of this revised gravitational theory include torsion, contorsion, and super-potential, which are expressed as follows \cite{Harko}
\begin{equation}\label{7a}
    {T^\lambda}_{\mu\nu}={g_a}^\lambda(\partial_\mu{g^a}_\nu-\partial_\nu{g^a}_\mu),
\end{equation}
\begin{equation}\label{7b}
    {K^{\mu\nu}}_\lambda=\frac{1}{2}({T_\lambda}^{\mu\nu}+{T^{\nu\mu}}_\lambda-{T^{\mu\nu}}_\lambda),
\end{equation}
\begin{equation}\label{7c}
    {S_\lambda}^{\mu\nu}=\frac{1}{2}({K^{\mu\nu}}_\lambda+{\delta^\mu}_\lambda {T^{\gamma\nu}}_\gamma-{\delta^\nu}_\lambda {T^{\gamma\mu}}_\gamma).
\end{equation}
The torsion scalar can be interpreted as
\begin{equation}\label{7d}
    T={T^\lambda}_{\mu\nu}{S_\lambda}^{\mu\nu}.
\end{equation}
The torsion scalar encapsulates the non-Riemannian characteristics of spacetime geometry in the presence of torsion. It serves as a quantitative indicator of the extent to which spacetime curvature diverges from the exclusively metric framework of GR.
The connection between $L_m$ and the energy-momentum tensor is articulated as follows
\begin{equation}\label{7}
\mathcal{T}_{ij}= - \frac{2}{\sqrt{-g}}\left[\frac{\partial(\sqrt{-g}L_{m})}{\partial g^{ij}}-\frac{\partial}{\partial x^{k}}\frac{\partial(\sqrt{-g}L_m)}{\partial(\partial g^{ij}/\partial x^{k})}\right].
\end{equation}
We consider that $L_m$ relies solely on the metric component and is not influenced by its derivatives
\begin{equation}\label{8}
\mathcal{T}_{ij}= - \frac{2}{\sqrt{-g}}\left[\frac{\partial(\sqrt{-g}L_{m})}{\partial g^{ij}}\right].
\end{equation}
By varying the action (\ref{6}) respect to the metric tensor, we derive the generalized field equations pertinent to this theory
\begin{eqnarray}\label{9}
    \left[e^{-1}\partial_\theta(ee^{\mu}_{a}S^{\nu\theta}_\mu)+e^\mu_aT^\theta_{\zeta\mu}S^{\zeta\nu}_\theta\right]f_T\nonumber \\
   + e^\mu_aS^{\nu\theta}_\alpha(f_{TT}\partial_\theta T+f_{T\mathcal{T}}\partial_\theta\mathcal{T})\nonumber \\
    +\frac{e^\nu_af}{4}\nonumber \\
    -\left(\frac{e^\mu_a\mathcal{T}^\nu_\mu+p_te^\nu_a}{2}\right)f_\mathcal{T}
    =\frac{e^\mu_a\,{T_\mu}^\nu}{4}.
\end{eqnarray}
where $f_T \equiv \frac{\partial f(T,\mathcal{T})}{\partial T}$, $f_\mathcal{T} \equiv \frac{\partial f(T,\mathcal{T})}{\partial \mathcal{T}}$ 

We consider a linear form for $f(T,\mathcal{T})$ as follow
\begin{equation}\label{9a}
    f(T, \mathcal{T})=\alpha T+\beta\mathcal{T}.
\end{equation}
where $\alpha$ is a constant and $\beta$ is the coupling parameter. Using (\ref{1}), one can find that
\begin{equation}\label{9b}
    T=-\frac{2}{r}(2\phi'+\frac{1}{r})(1-\frac{b}{r}).
\end{equation}
It is straightforward to show that metric (\ref{1}) and eq. (\ref{9})  give
\begin{equation}\label{f1}
    \rho(r)=Y(\alpha,\beta)\frac{(\beta-4)r\,b'(r)+\beta\,b(r)}{4(\beta-1)r^3},
\end{equation}
\begin{equation}\label{f2}
    p(r)=Y(\alpha,\beta)\frac{3\beta r\,b'(r)+(4-5\beta)\,b(r)}{4(\beta-1)r^3},
\end{equation}
\begin{equation}\label{f3}
    p_t(r)=Y(\alpha,\beta)\frac{(\beta+2)r\,b'(r)+(\beta-2)\,b(r)}{4(\beta-1)r^3}.
\end{equation}
which the prime denotes the derivative $\frac{d}{dr}$ and
 \begin{equation}\label{10}
Y(\alpha, \beta)=\frac{\alpha}{\beta+2}.
\end{equation}

Wormhole solutions within the framework of standard GR have been shown to violate ECs. To maintain a positive stress-energy tensor in the presence of matter, these ECs offer practical methodologies. The ECs, which encompass the NEC, dominant energy condition (DEC), weak energy condition (WEC), and strong energy condition (SEC), are explicitly defined to aid in achieving this goal,
\begin{eqnarray}\label{21}
\textbf{NEC}&:& \rho+p\geq 0,\quad \rho+p_t\geq 0 \\
\label{21a}
\textbf{WEC}&:& \rho\geq 0, \rho+p\geq 0,\quad \rho+p_t\geq 0, \\
\textbf{DEC}&:& \rho\geq 0, \rho-|p|\geq 0,\quad \rho-|p_t|\geq 0, \\
\textbf{SEC}&:& \rho+p\geq 0,\, \rho+p_t\geq 0,\rho+p+2p_t \geq 0. \label{21b}
\end{eqnarray}
The ECs fundamentally derive from the Raychaudhuri equation, which characterizes the dynamics of a congruence of curves that can be timelike, spacelike, or lightlike. The ECs possess numerous significant theoretical applications, utilized across various contexts to derive general conclusions that are applicable to a wide range of scenarios.
Employing these equations enables us to elucidate the intricate paths traversed by celestial bodies. At this juncture, as noted in \cite{fq}, we will examine the ECs in the forthcoming sections of this paper by defining the functions,
\begin{eqnarray}\label{22}
 H(r)&=& \rho+p ,\, H_1(r)= \rho+p_t,\, H_2(r)= \rho-|p|, \nonumber \\
 H_3(r)&=&\rho-|p_t|,\, H_4(r)= \rho+p+2p_t .
\end{eqnarray}
For simplicity, we will adopt the value of $r_0=1$ in the subsequent discussions within this document.

Let us study the ECs in the context of $f(T,\mathcal{T})$ gravity. It is easy to show that
\begin{equation}\label{24}
H(r)=\rho(r)+p(r)= Y(\alpha, \beta)\frac{rb'-b}{r^3}
\end{equation}
and
\begin{equation}\label{24b}
H_1(r)=Y(\alpha, \beta)\frac{rb'+b}{2r^3}.
\end{equation}
The flaring-out condition is crucial for the construction of a traversable wormhole, which allows one to conclude that
\begin{equation}\label{24c}
Y(\alpha, \beta)<0
\end{equation}
can result in $H>0$. In general, it is possible to demonstrate that
\begin{equation}\label{24ccc}
H(r)=Y(\alpha, \beta)H^{GR}, \qquad H_1(r)=Y(\alpha, \beta)H_{1}^{GR}
\end{equation}
where $H^{GR}$ and $H_{1}^{GR}$represent the radial and lateral NEC within the framework of standard GR. Consequently, we can infer that solutions which contravene both the radial and lateral NEC in the context of GR may still satisfy the NEC when considered within the background of $f(T,\mathcal{T})$ gravity, provided that condition (\ref{24c}) remains applicable. It is easy to show that conditions (\ref{24c}) is valid in the range
\begin{equation}\label{4cc}
\alpha<0,\, -2<\beta\quad or \quad \alpha>0,\, \beta<-2
\end{equation}

Researchers have employed various approaches to identify asymptotically flat wormhole solutions within the framework of GR. The most favored method involves the consideration of an EoS. Alternatively, some researchers utilize shape functions characterized by free parameters, subsequently fine-tuning these parameters in an effort to discover solutions that align with both physical and mathematical constraints. In the next section, we apply these methods to find asymptotically flat wormhole solutions.

\section{Non-exotic Wormhole solutions }\label{sec3}
\begin{figure}
\centering
  \includegraphics[width=3 in]{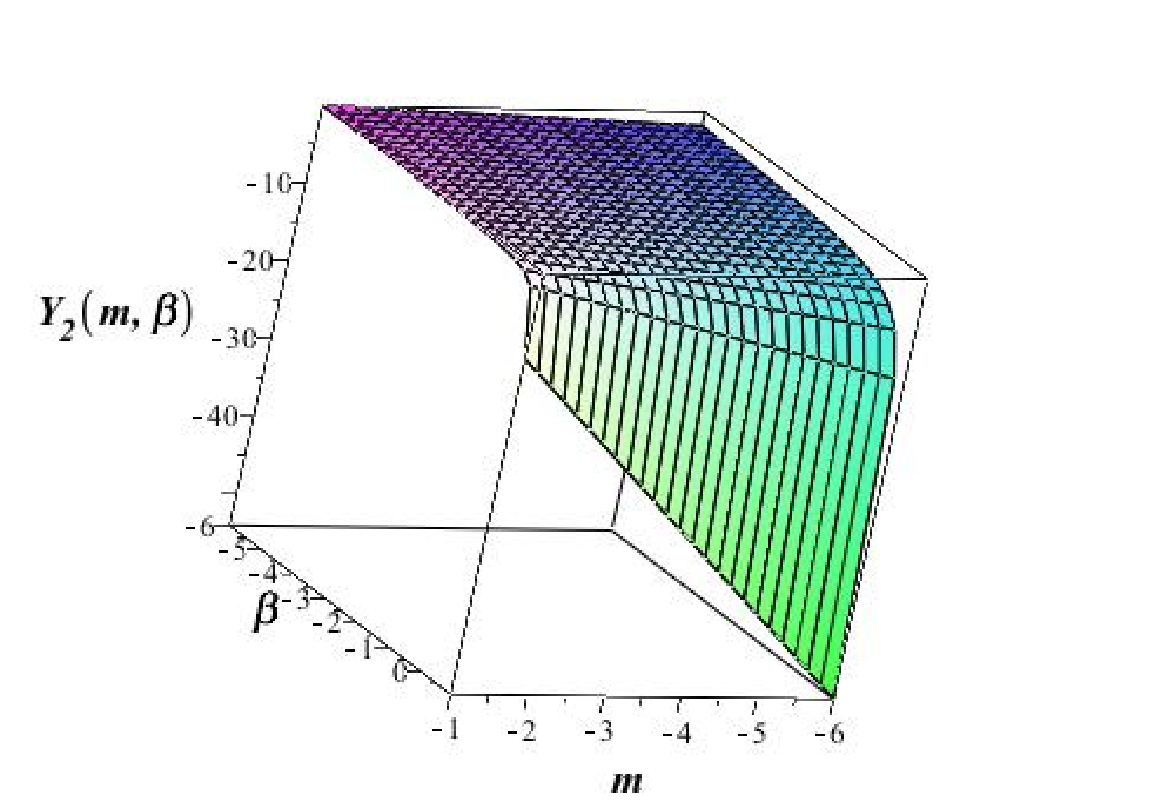}
\caption{The figure represents the $Y_2(m,\beta)$ against $m$ and $\beta$,  which is  negative in the entire range. See the text for details.}
 \label{fig1}
\end{figure}

\begin{figure}
\centering
  \includegraphics[width=3 in]{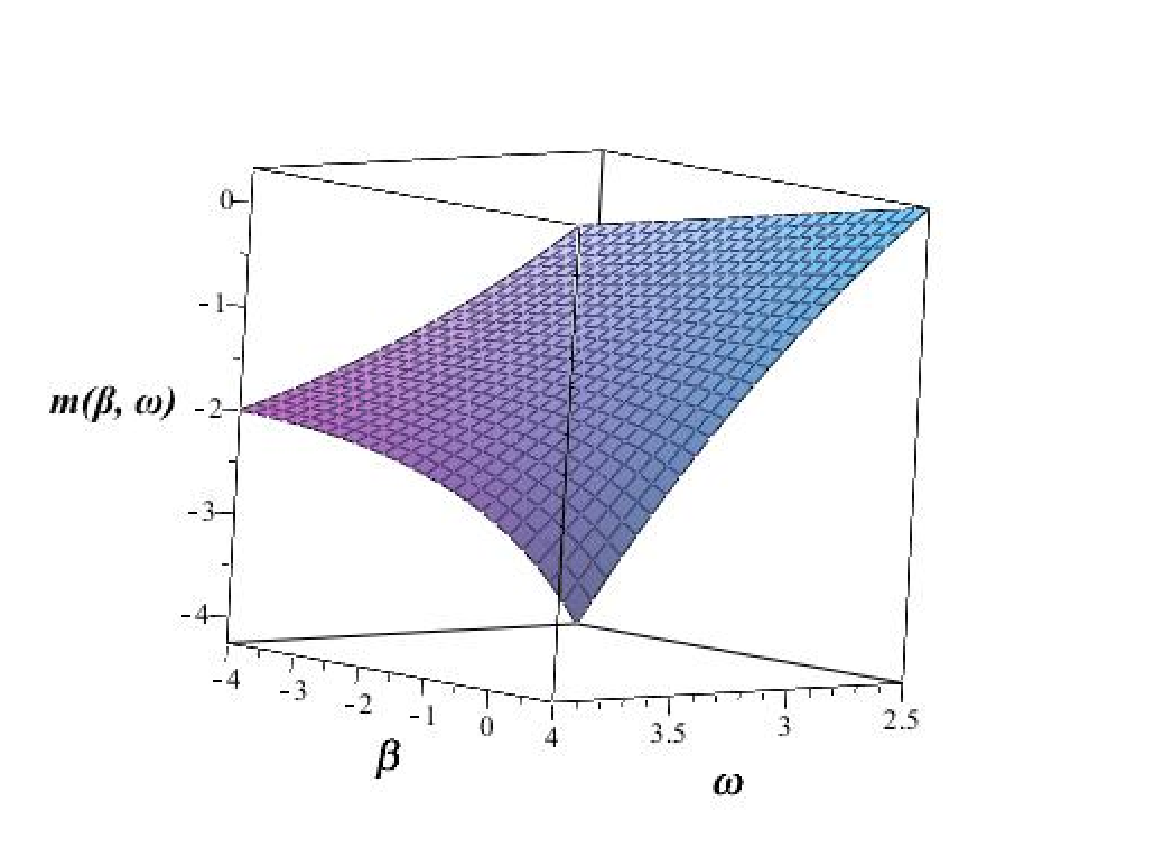}
\caption{The graph illustrates the relationship between $m(\omega,\beta)$ and the variables $\omega$ and $\beta$. It indicates that within a certain range, $m$ can take values less than -1, resulting in an asymptotically flat shape function. It is clear that the function $m(\omega,\beta)$ exhibits an increasing trend with respect to $\beta$ and a decreasing trend with respect to $\omega$. See the text for details.}
 \label{fig2}
\end{figure}

It was established that solutions which violate bout lateral and radial NEC in the background of GR can respect NEC in the context of $f(T,\mathcal{T})$ when condition (\ref{24c}) holds. In this section, we explore some of the potential wormhole solutions in this scenario. As the first example, the solution with a linear EoS as follows
\begin{equation}\label{24a}
p_r (r)=\omega\rho(r)
\end{equation}
is considered. Using(\ref{f1}), (\ref{f2}) and (\ref{24a}) gives
\begin{equation}\label{2a8}
\frac{b'}{b}=\frac{m}{r}
\end{equation}
where
\begin{equation}\label{27}
m=\frac{\beta(5+\omega)-4}{\beta(3-\omega)+4\omega}.
\end{equation}
Solving this equation leads to
\begin{equation}\label{241}
b(r)=r^m.
\end{equation}
The shape function (\ref{241}) is widely recognized within the context of wormhole theory \cite{fq, fq2}. It satisfies all essential criteria required for the existence of traversable wormhole solutions. Naturally, it is imperative to consider the asymptotically flat solution, necessitating the condition that $m<1$ be imposed. Let us investigate energy density and lateral pressure in this case. It is easy to show that
\begin{equation}\label{11}
\rho(r)=Y(\alpha,\beta)Y_2(m,\beta)r^{m-3}.
\end{equation}
where
\begin{equation}\label{13a}
Y_2(m,\beta)=\frac{m\beta-4m+\beta}{4(\beta-1)}.
\end{equation}
Since $Y<0$ is imposed,
\begin{equation}\label{13aa}
Y_2(m,\beta)<0
\end{equation}
is the essential condition to have a positive energy density. In the next step, we analyze the lateral  NEC. It is easy to show that
 \begin{equation}\label{241a}
H_1(r)=\rho+p_t=\frac{m+1}{2}Y(\alpha,\beta)r^{m-3}.
\end{equation}
Using (\ref{24c}) implies that
\begin{equation}\label{241b}
m<-1.
\end{equation}
must be held to satisfy the lateral NEC. One can show that
 \begin{equation}\label{24cc}
H_4(r)=Y(\alpha,\beta)(3m-1)\frac{\beta}{2(\beta-1)}r^{m-3}.
\end{equation}
Using (\ref{24c}), (\ref{241b}) and (\ref{24cc}) explains that $H_4>0$ is valid for the whole range of $\beta$ except
 \begin{equation}\label{2c}
0<\beta<1.
\end{equation}
We have plotted $Y_2(m,\beta)$ as a function of $m$ and $\beta$ in Fig.(\ref{fig1}) which is negative in the entire range $\beta<1$ and $m<-1$. It can be shown that for $\beta>1$ and $m<-1$, the sign of $Y_2(m,\beta)$ is positive.
 Figure (\ref{fig1}) guaranties that condition (\ref{13aa}) is valid for the range $\beta<1$ and $m<-1$. It is clear that the function $Y_2(m,\beta)$  exhibits a decreasing trend with respect to $m$.
 Figure (\ref{fig2}) represent $m(\omega,\beta)$ as a function of $\omega$ and $\beta$ which shows that $-1>m$ can be valid in some range of $\omega$ and $\beta$. It is evident that $m(\omega,\beta)$ increases as $\beta$ ascends and decreases as $\omega$ ascends.

\begin{figure}
\centering
  \includegraphics[width=3 in]{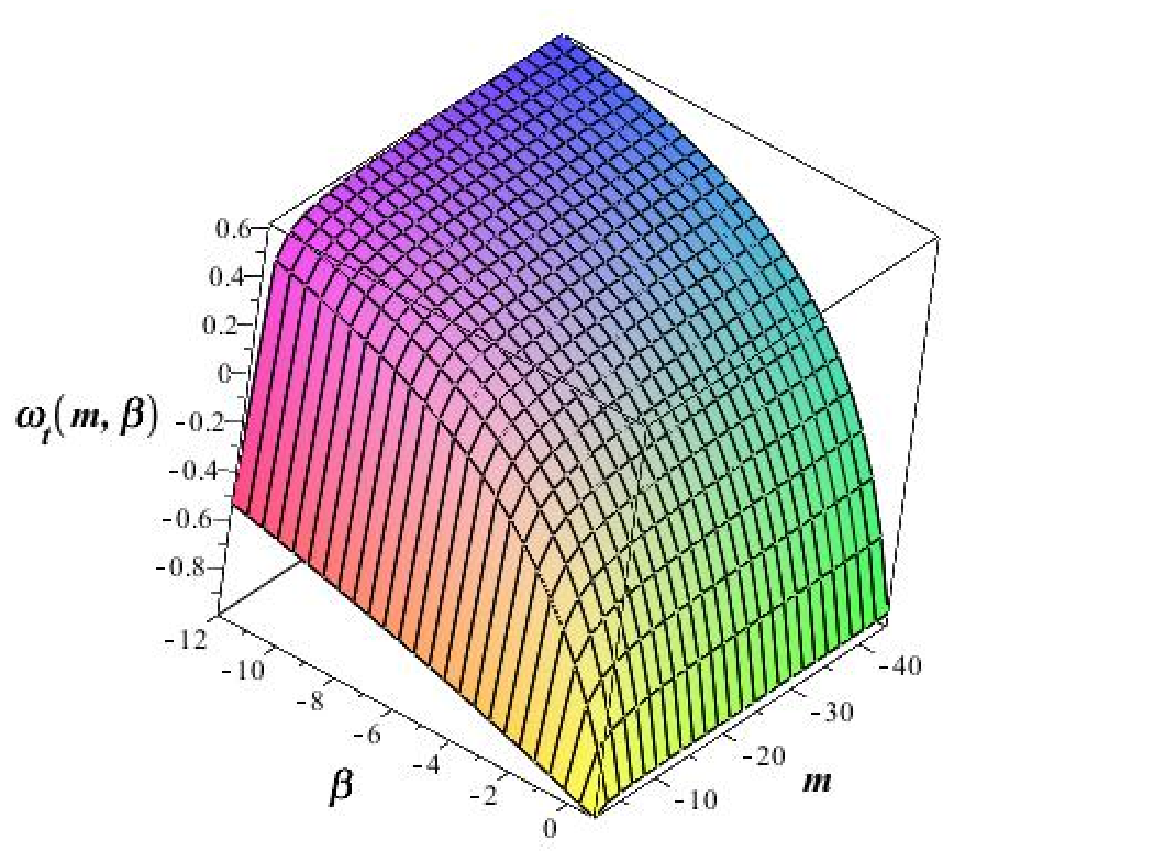}
\caption{The plot depicts $\omega_t(r)$ against $m$ and $\beta$ which shows the lateral EoS parameter is in the range $-1<\omega_t<1$. See the text for details.}
 \label{fig3}
\end{figure}

It is easy to show that
\begin{equation}\label{24cd}
\omega(m,\beta)=\frac{p(r)}{\rho(r)}=\frac{\beta(3m-5)+4}{m(\beta-4)+\beta}
\end{equation}
and
\begin{equation}\label{24fc}
\omega_t(m,\beta)=\frac{p_t(r)}{\rho(r)}=\frac{m(\beta+2)+\beta-2}{m(\beta-4)+\beta}.
\end{equation}
We have plotted $\omega_t(m,\beta)$ against $m$ and $\beta$ in the range $-1>m$ and $\beta<1$ in Fig.(\ref{fig3}) which shows
\begin{equation}\label{24d}
 -1<\omega_t<1.
\end{equation}

\begin{figure}
\centering
  \includegraphics[width= 3 in]{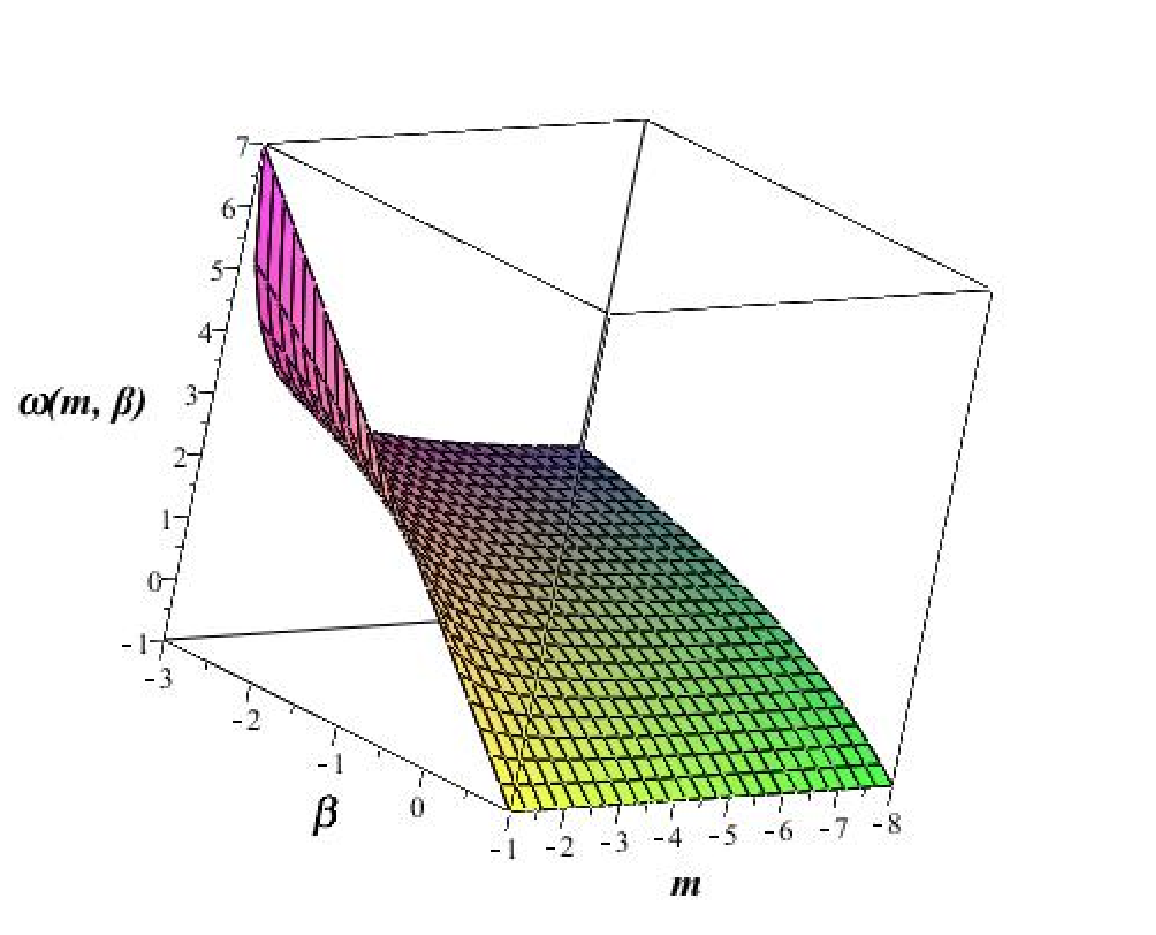}
\caption{The figure represents the $\omega(m, \beta)$ against $m$ and $\beta$ ,  which shows $-1<\omega$ in the entire range for $m<-1$ and $\beta<1$ . See the text for details.}
 \label{fig4}
\end{figure}

Also, the function $\omega(m,\beta)$ has been graphed with respect to the variables $m$ and $\beta$ within the intervals $-1 > m$ and $ \beta < 1$, as illustrated in  Fig. (\ref{fig4}). This figure indicates that
\begin{equation}\label{24d1}
-1 <w
\end{equation}
is accessible in these ranges of $m$ and $\beta$. Given that $\rho$ is positive, Eq. (\ref{24d}) ensures that the condition $H_3>0$ holds true. Furthermore, it can be inferred that $H_2>0$ is also valid for the interval $-1 < \omega < 1$. In conclusion, it can be asserted that the NEC, WEC, and SEC for the shape function $b(r)=r^m$ are satisfied within the ranges $m<-1$, and $\beta<0$. The DEC is satisfied when the condition $-1 < \omega < 1$ is met.

\subsection{ isotropic wormhole solutions }\label{subsec1}

\begin{figure}
\centering
  \includegraphics[width=3 in]{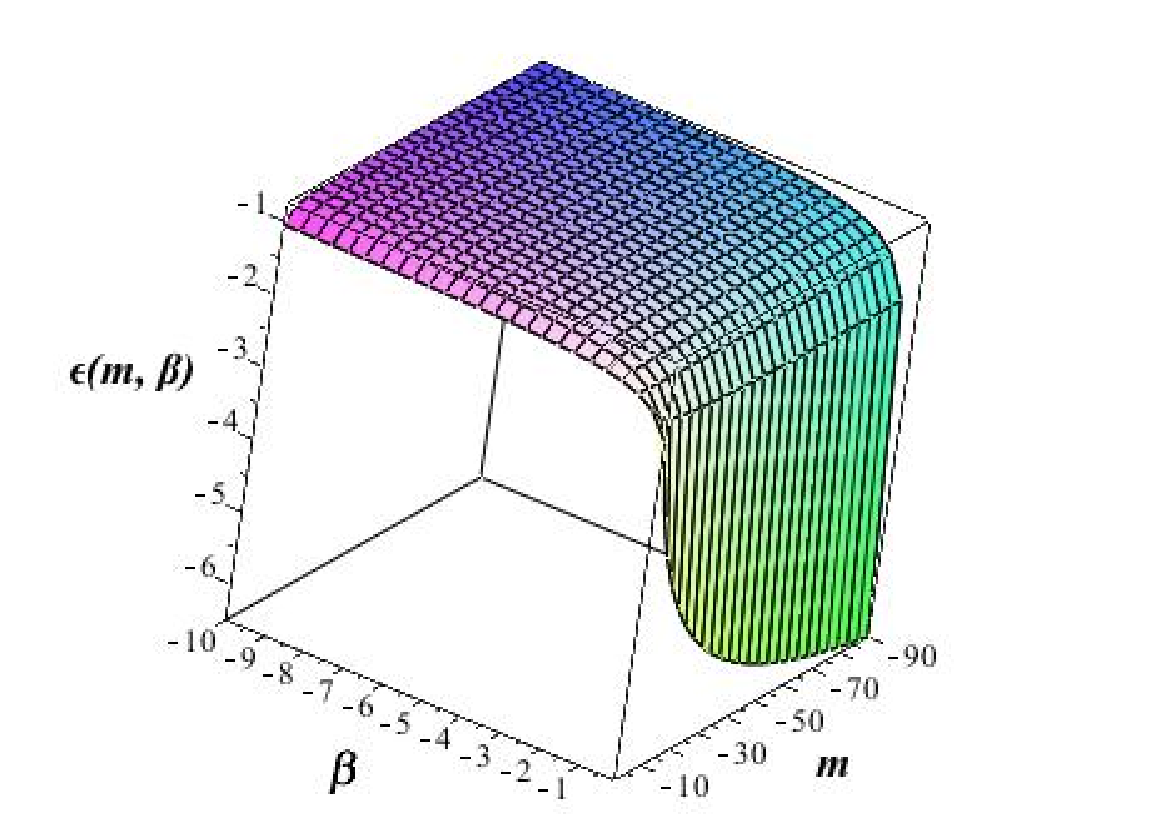}
\caption{The plot depicts $\epsilon(m, \beta)$ against $m$ and $\beta$,  showing a non-vanishing behaviour in the entire range. See the text for details.}
 \label{fig5}
\end{figure}

In this subsection, we study the isotropic wormhole solutions in the context of $f(T,\mathcal{T})$ for a power-law shape function. It is easy to show that
\begin{eqnarray}\label{25}
 \omega=\omega_t.
\end{eqnarray}
leads to $m=3$ which does not satisfy the asymptotically flat condition. In the next step, we try to find solutions which are approximately isotropic. To achieve these solutions, we define a parameter as follows
 \begin{equation}\label{26}
\epsilon(m, \beta)=\frac{\omega_t}{\omega}-1,
\end{equation}
This parameter can be interpreted as a scale for isotopic characteristics of a wormhole. One can see that
\begin{equation}\label{27a}
\epsilon(m, \beta)=\frac{-2m\beta+2m+6\beta-6}{3m\beta-5\beta+4}
\end{equation}
We have plotted $\epsilon(m, \beta)$ as a function of $m$ and $\beta$ in the range $m<-1$ and $\beta<0$ in Fig.(\ref{fig5}). This figure implies that the $\epsilon$ is not vanishing in the entire range. We can therefore conclude that the nearly isotropic solutions are not attainable within the range where the solutions satisfy the ECs.

\subsection{ Solutions for $p(r)=n p_t(r)$ }\label{subsec2}
\begin{figure}
\subfloat[$0<n$]{\includegraphics[width = 2.2 in]{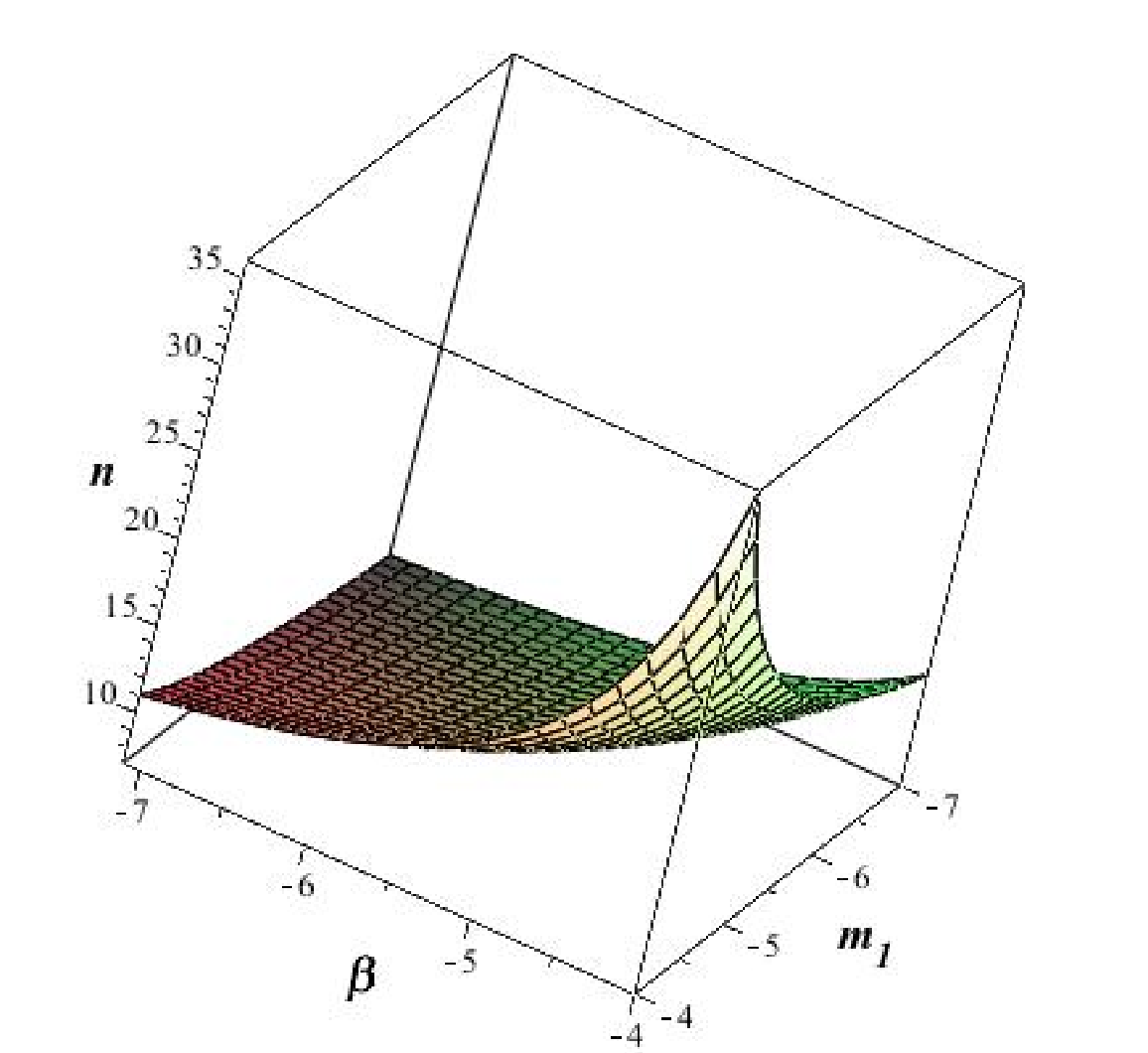}}\\
\subfloat[$n<0$]{\includegraphics[width = 2.2 in]{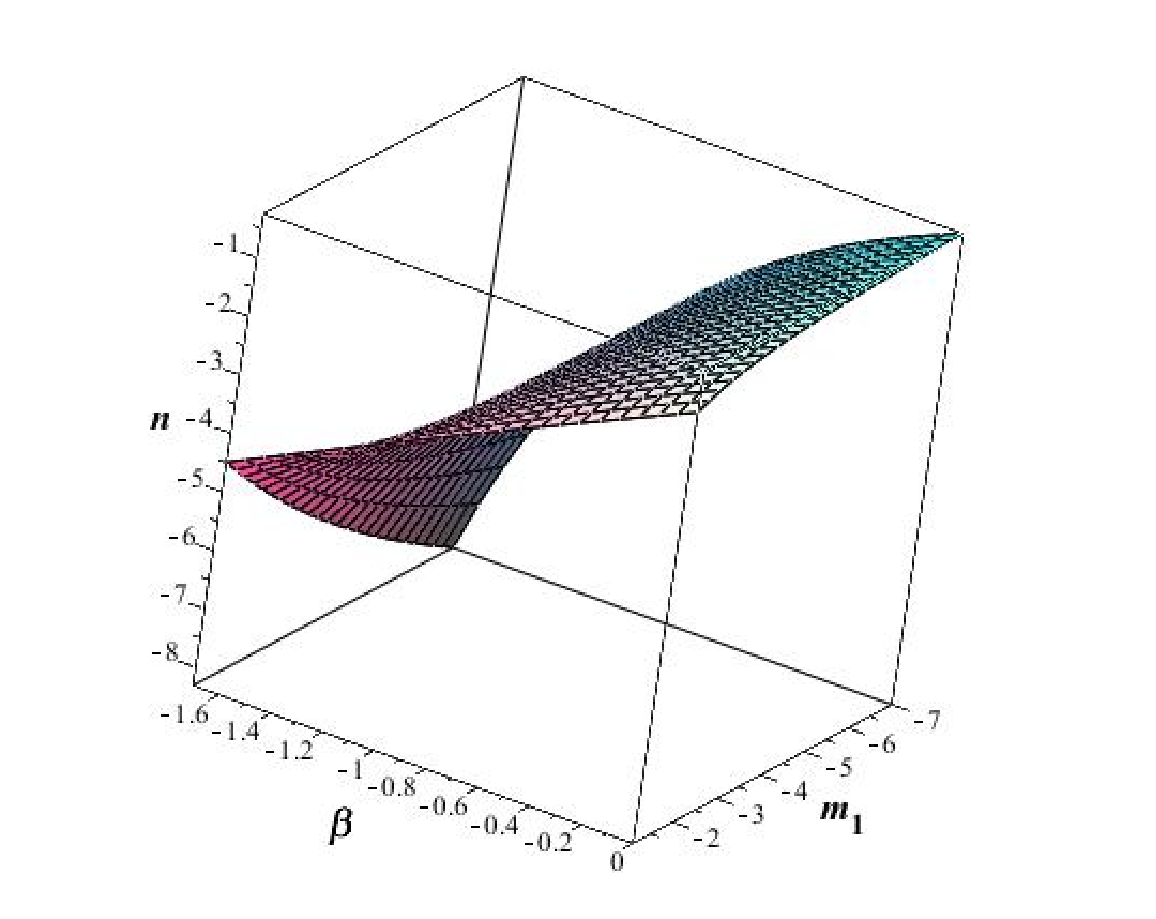}}
\caption{The graph depicts the correlation between $n(m_1, \beta)$ and the variables $m_1$ and $\beta$. It is clear that $n$ is positive within a certain range of $m_1$ and $\beta$ (a), while $n$ is negative in a different range (b).}
\label{fig6}
\end{figure}

In this subsection, we chose an EoS as follows
\begin{equation}\label{28bb}
p(r)=np_t(r).
\end{equation}
Using Eqs.(\ref{f2}) and (\ref{f3}) with (\ref{28bb}) gives
\begin{equation}\label{29}
b(r)=r^{m_1}.
\end{equation}
where
\begin{equation}\label{a29}
m_1=\frac{4-5\beta-n\beta+2n}{n\beta+2n-3\beta}.
\end{equation}
It is clear that this class of solutions admit the linear EoS. In other words, solutions with linear EoS correspond to solutions with EoS in the form (\ref{28bb}) where
\begin{equation}\label{a291}
n=\frac{\omega}{\omega_t}=\frac{\beta(3m_1-5)+4}{(\beta+2)m_1+\beta-2}.
\end{equation}
The general behavior of $n$ as a function of $\beta$ and $m_1$ can be seen in Fig.(\ref{fig6}). This figure illustrates that lateral pressure can either have the same sign as radial pressure or an opposite sign.

\subsection{ Solutions for $\mathcal{T}=0$ }\label{subsec3}
Now, we investigate wormhole solutions with $\mathcal{T}=0$ in the context of $f(T,\mathcal{T})$ . In this case, EoS is
\begin{equation}\label{38}
\mathcal{T}=-\rho+p+2p_t=0.
\end{equation}
which leads to
\begin{equation}\label{39}
b(r)=r^{\frac{\beta}{\beta+2}}.
\end{equation}
This is a special case of (\ref{241}) with $m=\frac{\beta}{\beta+2}$. The asymptotically flat conditions permit the value of $\beta$ to fall within the range of
 \begin{equation}\label{39a}
\beta>-2.
\end{equation}
Using $m<-1$ gives
\begin{equation}\label{39b}
\beta<-1.
\end{equation}
Equations (\ref{39a}) and (\ref{39b}) allow $\beta$ to be in the range $-2<\beta<-1$.
In this case, the EoS parameters are as follow
\begin{eqnarray}\label{a22}
 \omega(\beta)=\frac{p}{\rho}=-\frac{\beta+4}{\beta}, \nonumber \\
  \omega_t(\beta)=\frac{p_t}{\rho}=\frac{\beta+2}{\beta}.
\end{eqnarray}
Using this equation in the range $-2<\beta<-1$ gives
\begin{eqnarray}\label{a39}
1<\omega<3, \nonumber \\
 -1< \omega_t<0.
\end{eqnarray}
Equation (\ref{a39}) explains that $H_2>0$ can not be valid for this category of solutions so the DEC is not valid for this case. Given that the energy density is positive, Equation (\ref{a39}) suggests that the radial pressure is also positive, while the lateral pressure is negative for this category of solutions. It is easy to show that this class of solutions admits all of the ECs except DEC.

\section{Solutions with variable EoS}\label{sec4}
It is important to note that solutions that contravene both radial and lateral NEC within the framework of GR are not readily accessible. As demonstrated in \cite{SR1}, a considerable proportion of the shape functions documented in the literature fail to satisfy these two conditions concurrently. In this section, we employ a variable EoS to identify solutions that violate both radial and lateral NEC in the context of GR. Solutions with variable EoS can play a significant role in wormhole theory. We have presented some classes of solutions with variable EoS in \cite{variable} in the context of GR which minimizes the usage of exotic matter. Variable EoS may be perceived as more physically relevant, as the linear EoS functions as a universal equation; however, at a localized scale, strict adherence to a linear model is unnecessary \cite{variable, SR1}. This notion can be associated with the distinctive geometry of a wormhole near its throat. The introduction of a variable EoS parameter has revitalized the investigation of wormhole physics in contrast to a constant parameter. It diminishes the requirement for exotic matter within the framework of GR \cite{variable} and facilitates the presentation of a broad spectrum of non-exotic solutions in the context of $f(R,T)$ \cite{SR1}. In the following section, we use this method to find asymptotically flat wormhole solutions in the context of $f(T,\mathcal{T})$.
According to \cite{variable}, we consider a linear-like EoS as follows
\begin{equation}\label{f30}
p=\omega_{eff}(r)\rho(r)=(\omega_\infty+g(r))\rho(r).
\end{equation}
The term $\omega_{eff}(r)$ represents the effective state parameter, while $\omega_\infty$ signifies the constant state parameter at a significantly large radial coordinate. To achieve an asymptotically linear EoS, we establish the requisite condition
\begin{equation}\label{f31}
\lim_{r\rightarrow \infty}g(r)=0.
\end{equation}
Using Eqs.(\ref{f1}-\ref{f2}) and (\ref{f30}), gives
\begin{equation}\label{8b1}
b(r)=C \exp\left(\int\frac{-\omega_\infty\beta-g(r)\beta+4-5\beta}{r(\omega_\infty\beta-4\omega_\infty+g(r)\beta-4g(r)-3\beta)}dr\right),
\end{equation}
where $C$  can be found through condition (\ref{2}). In the next subsections, we use this equation for some different $g(r)$ functions to find desired wormhole solutions.

\subsection{ Solution for $g(r)=\frac{D}{r}$ }\label{subsec5}
\begin{figure}
\centering
  \includegraphics[width=3 in]{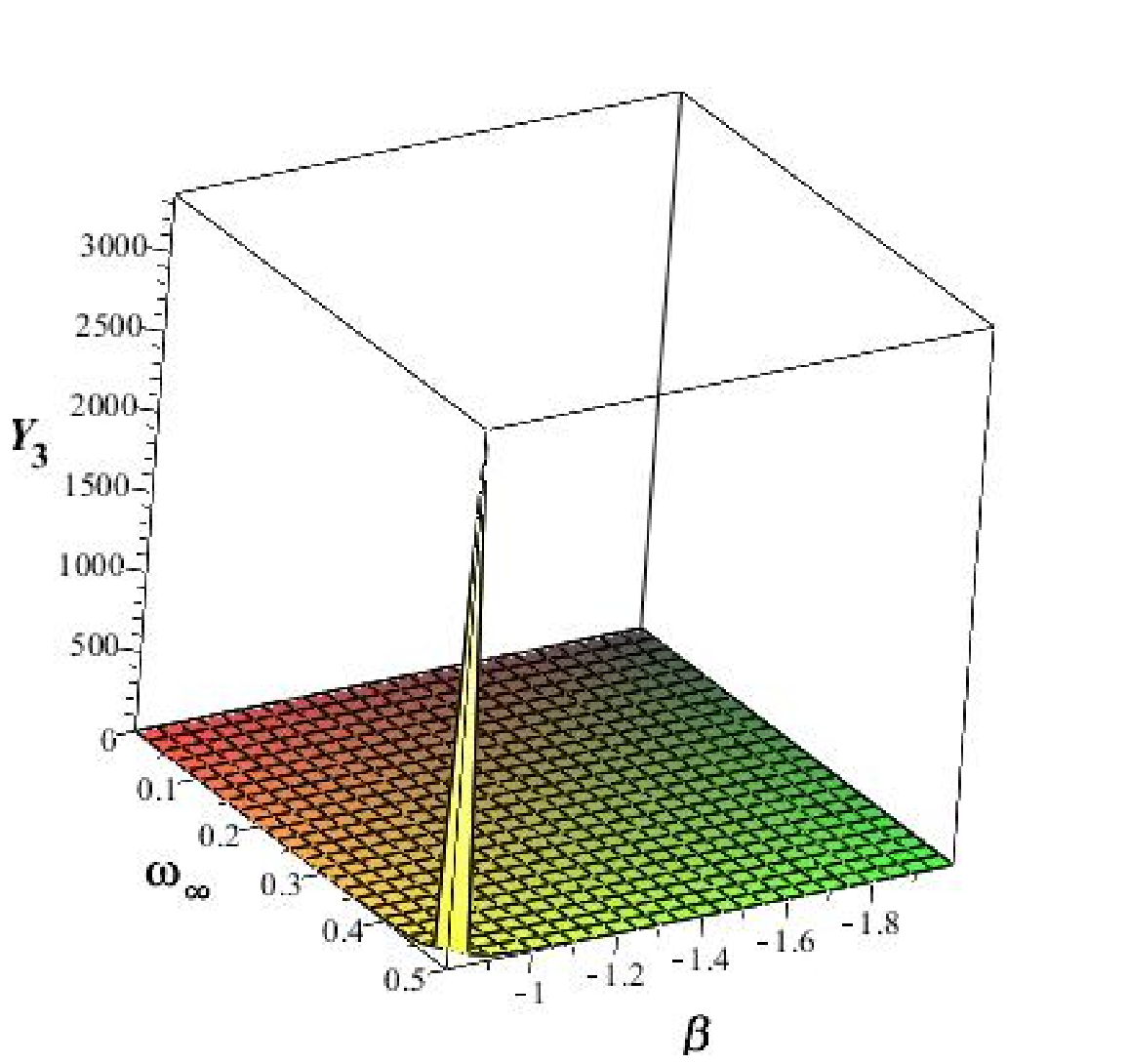}
\caption{The graph illustrates $Y_3(\omega_\infty,\beta)$ as a function of $\omega_\infty$ and $\beta$. It is observed to be positive within the range of $0 \leq \omega_\infty \leq 1/2$ and $-0.8 < \beta $. Consequently, the shape function  exhibits asymptotic flatness in these specified regions. See the text for details}
 \label{fig7}
\end{figure}

\begin{figure}
\centering
  \includegraphics[width=3 in]{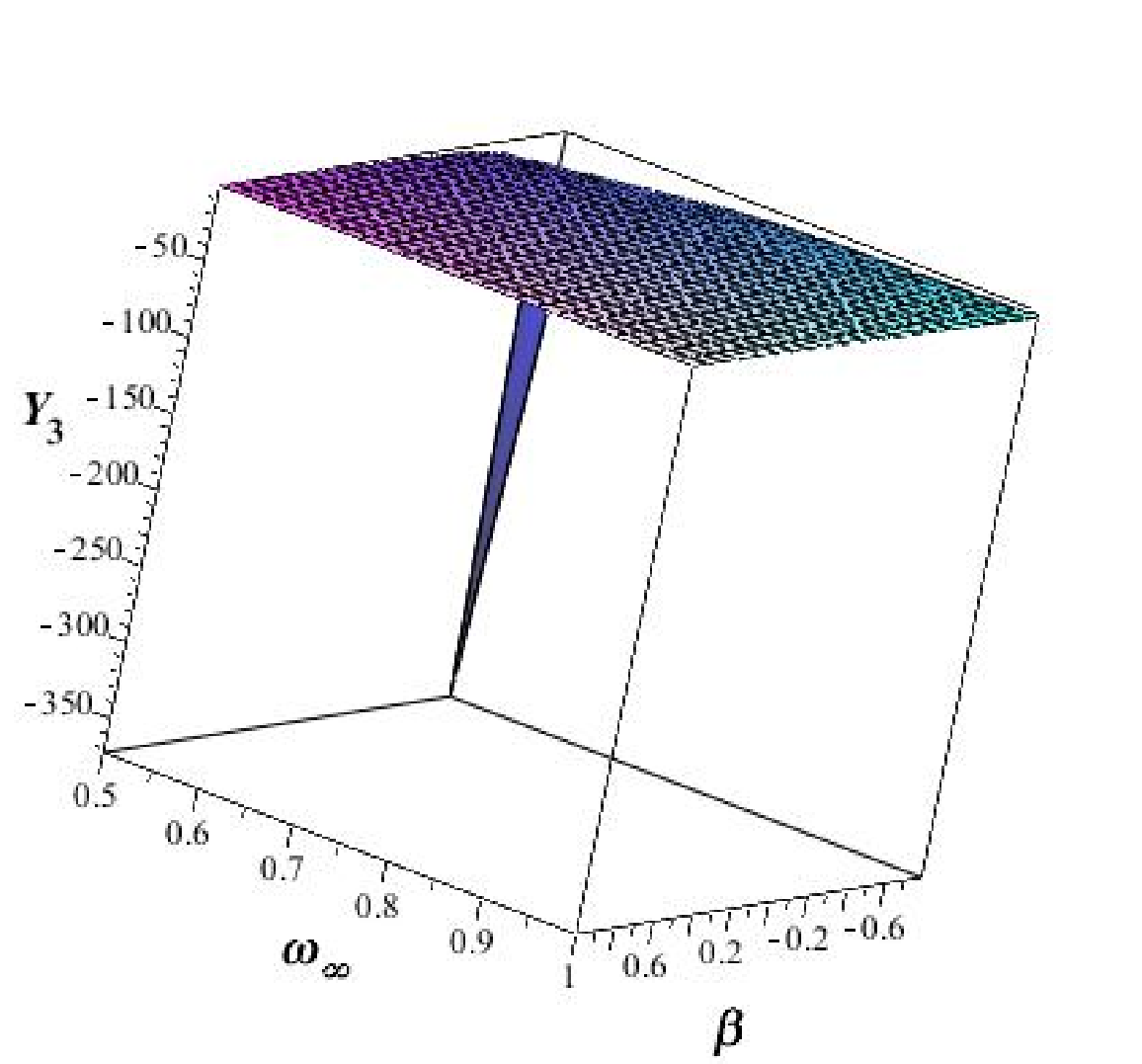}
\caption{The graph illustrates $Y_3(\omega_\infty,\beta)$ as a function of $\omega_\infty$ and $\beta$. It is observed to be negative within the range of $1/2 \leq \omega_\infty \leq 1$ and $ \beta > -0.8$. Consequently, the shape function does not exhibit asymptotic flatness in these specified regions. See the text for details}
 \label{fig8}
\end{figure}

One of the most straightforward options for $g(r)$ is
\begin{equation}\label{30}
g(r)=\frac{D}{r}.
\end{equation}
Using this function leads to
\begin{eqnarray}\label{31}
b(r)&=&C\, r^{-\frac{\beta}{\beta-4}}\nonumber \\
&\times&\left((\omega_\infty\beta-4\omega_\infty-3\beta)r+D\beta-4D\right)^{\gamma(\omega_\infty,\beta)}
\end{eqnarray}
where
\begin{equation}\label{32}
\gamma(\omega_\infty,\beta)=-\frac{8(\beta^2-3\beta+2)}{(\beta-4)(\omega_\infty\beta-4\omega_\infty-3\beta)}.
\end{equation}
To check asymptotically flat condition, we define
\begin{equation}\label{33}
Y_3(\omega_\infty,\beta)=\gamma(\omega_\infty,\beta)-\frac{\beta}{\beta-4}-1.
\end{equation}
It is clear that the condition
\begin{equation}\label{34}
Y_3(\omega_\infty,\beta)<0
\end{equation}
must be satisfied to have an asymptotically flat solution.
We have plotted $Y_3(\omega_\infty,\beta)$ as a function of $\omega_\infty$ and $\beta$ in two different regions for $\beta$ in Figs. (\ref{fig7}) and (\ref{fig8}). These figures indicate that condition (\ref{34}) is satisfied in some regions of $\omega_\infty$ and $\beta$. The behavior of the shape function is influenced by the parameters $\omega_\infty$ and $\beta$. To further investigate the properties of the solutions, we will examine specific cases. As the first model, we study the wormhole solutions for $\omega_\infty=1/2$ which leads to
\begin{eqnarray}\label{34b}
b(r)&=&(2D\beta-8D-5\beta r-4r)^{\gamma(1/2,\beta)}\nonumber \\
 &\times& r^{-\frac{\beta}{\beta-4}}(2D\beta-8D-5\beta -4)^{-\gamma(1/2,\beta)}.
\end{eqnarray}
In this case, $Y_3(1/2,\beta)$ is
\begin{equation}\label{35}
Y_3(1/2,\beta)=\frac{11\beta-8}{5\beta+4}.
\end{equation}
One can deduce that the asymptotic solution is attainable within the interval $-0.8<\beta<8/11$. We can now proceed to examine the ECs. Analyzing the ECs in their general form for the free parameters $D$ and $\beta$ presents challenges. To simplify our analysis, we will consider the special case where $\beta=-1/2$. In this scenario, the shape function is as follows
\begin{equation}\label{37}
b(r)=\frac{(3/2+9D)^{80/9}}{(3r/2+9D)^{80/9}r^{1/9}}.
\end{equation}


\begin{figure}
\subfloat[$H_2>0$]{\includegraphics[width = 2.4in]{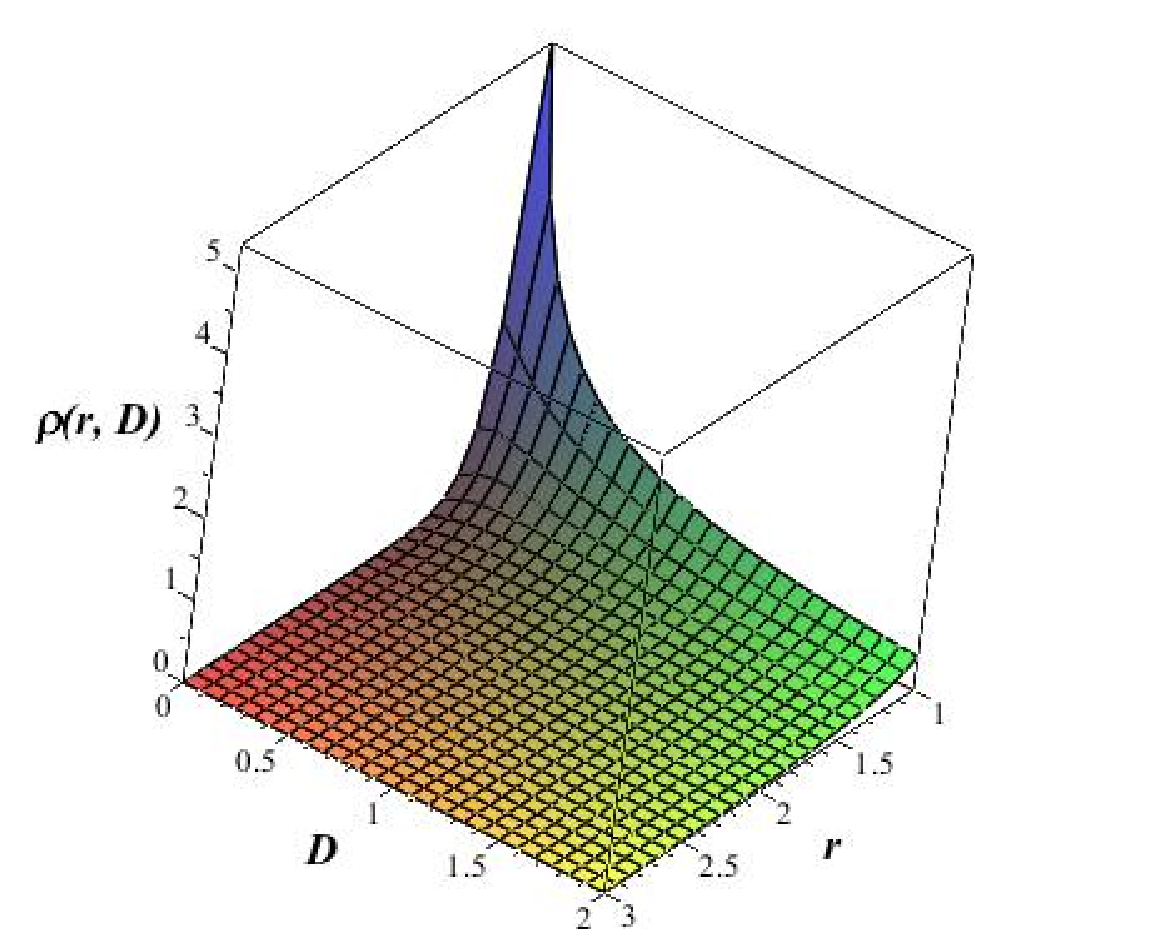}}\\
\subfloat[$H_4>0$]{\includegraphics[width = 2.4in]{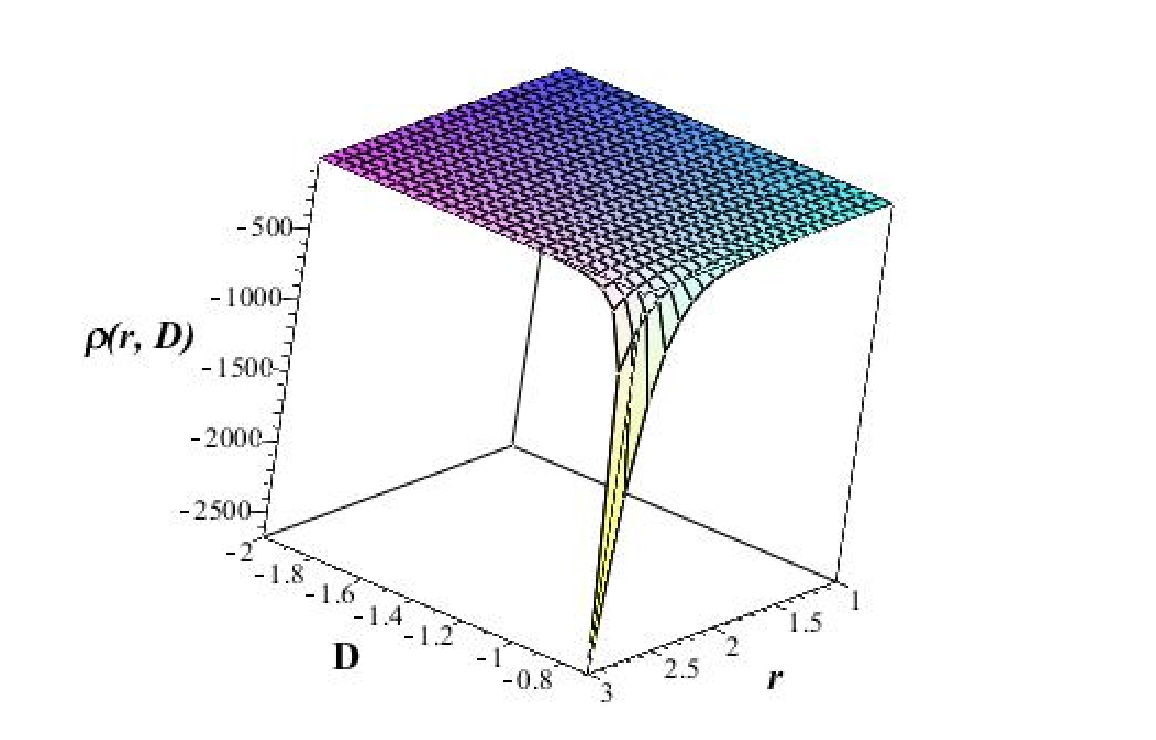}}
\caption{Energy density as a function of $r$ and $D$ for $\alpha=\beta=-1/2$. It is clear that $\rho$ is positive in the range $D>0$(a), while $\rho$ is negative  in the range $D<-0.7$ (b).
See the text for details.}
\label{fig9}
\end{figure}

\begin{figure}
\centering
  \includegraphics[width=3 in]{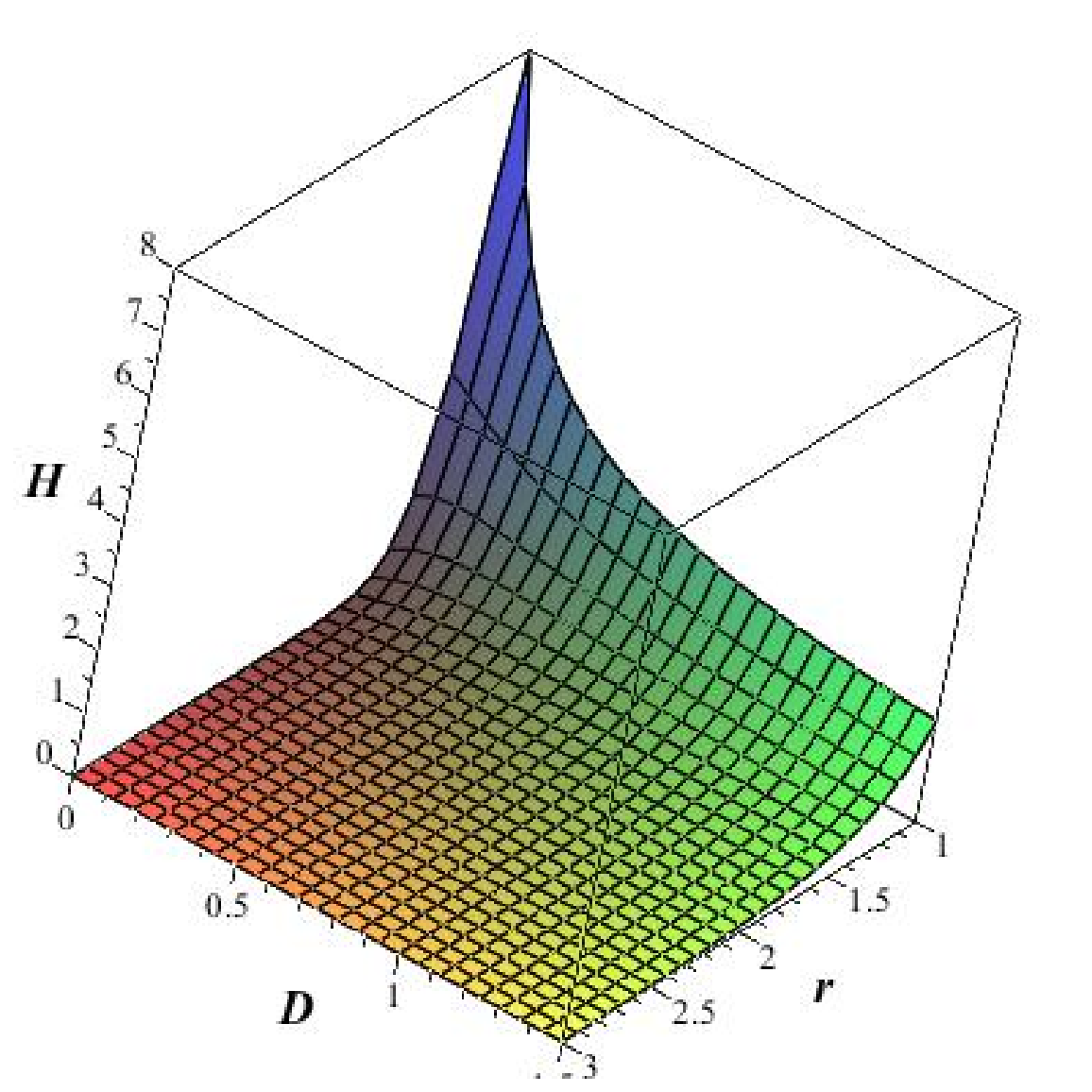}
\caption{The plot depicts $H(r,D)$ as a function of $r$ and $D$for $\alpha=\beta=-1/2$, which is positive for $D>0$. See the text for details.}
 \label{fig10}
\end{figure}
\begin{figure}
\centering
  \includegraphics[width=3 in]{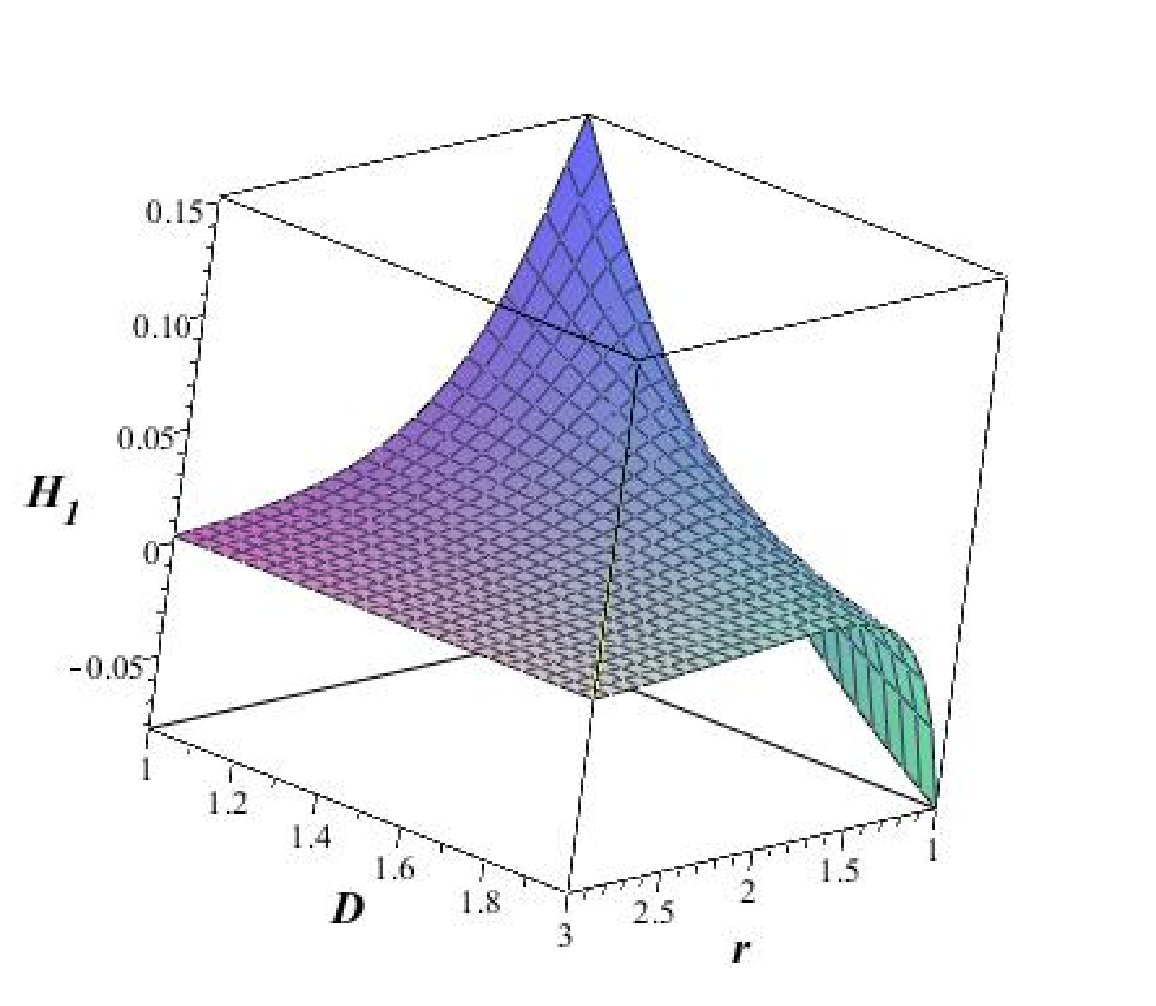}
\caption{The plot depicts $H_1(r,D)$ as a function of $r$ and $D$ for $\alpha=\beta=-1/2$ which is positive for some range and negative in the other range of $D>0$. See the text for details.}
 \label{fig11}
\end{figure}
\begin{figure}
\centering
  \includegraphics[width=3 in]{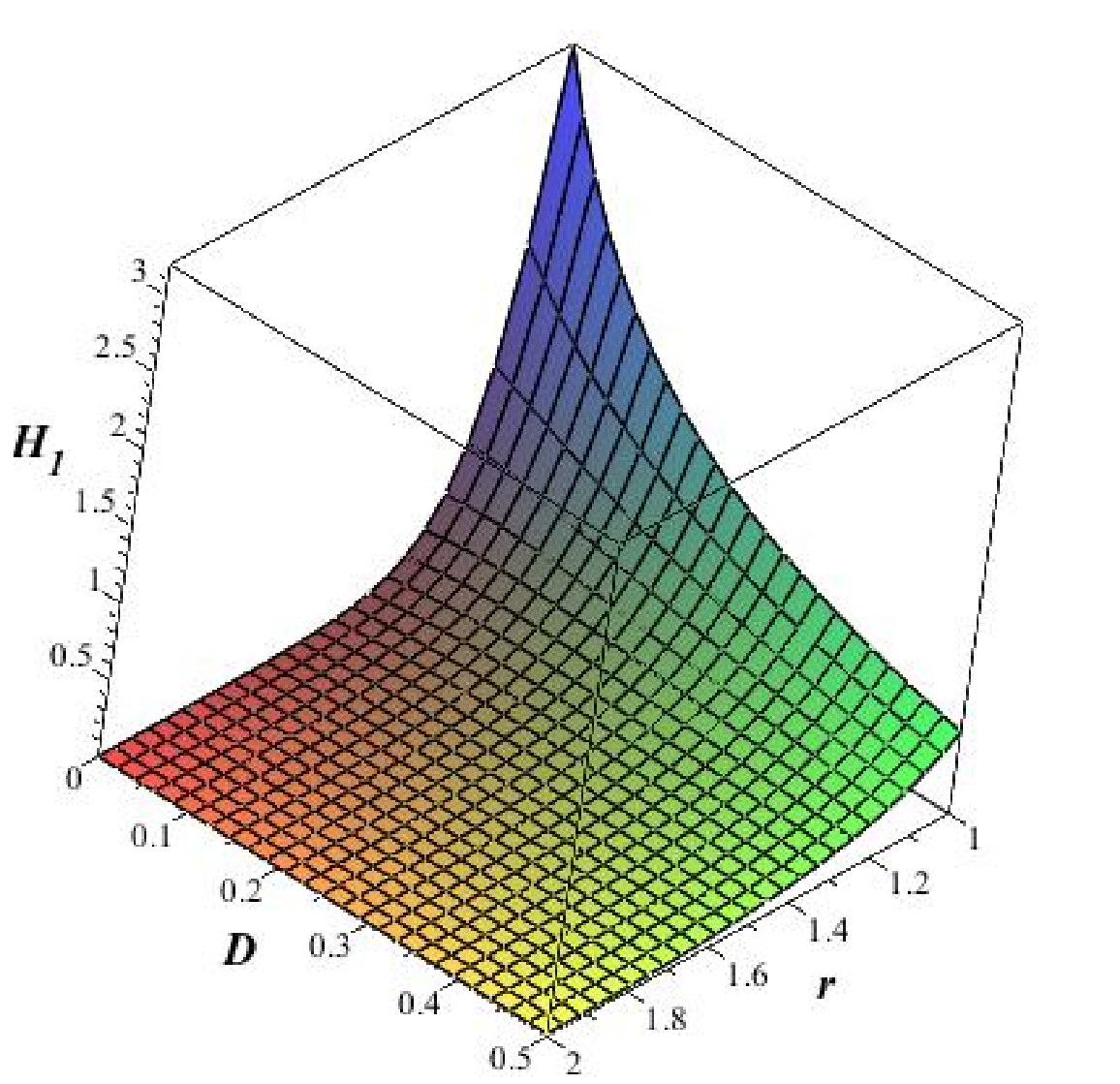}
\caption{$H_1(r,D)$ as a function of $r$ and $D$ for $\alpha=\beta=-1/2$. It is clear that $H_2$ is positive in the range $0<D< 1/2$. See the text for details}
 \label{fig12}
\end{figure}
\begin{figure}
\subfloat[$H_2>0$]{\includegraphics[width = 2.5in]{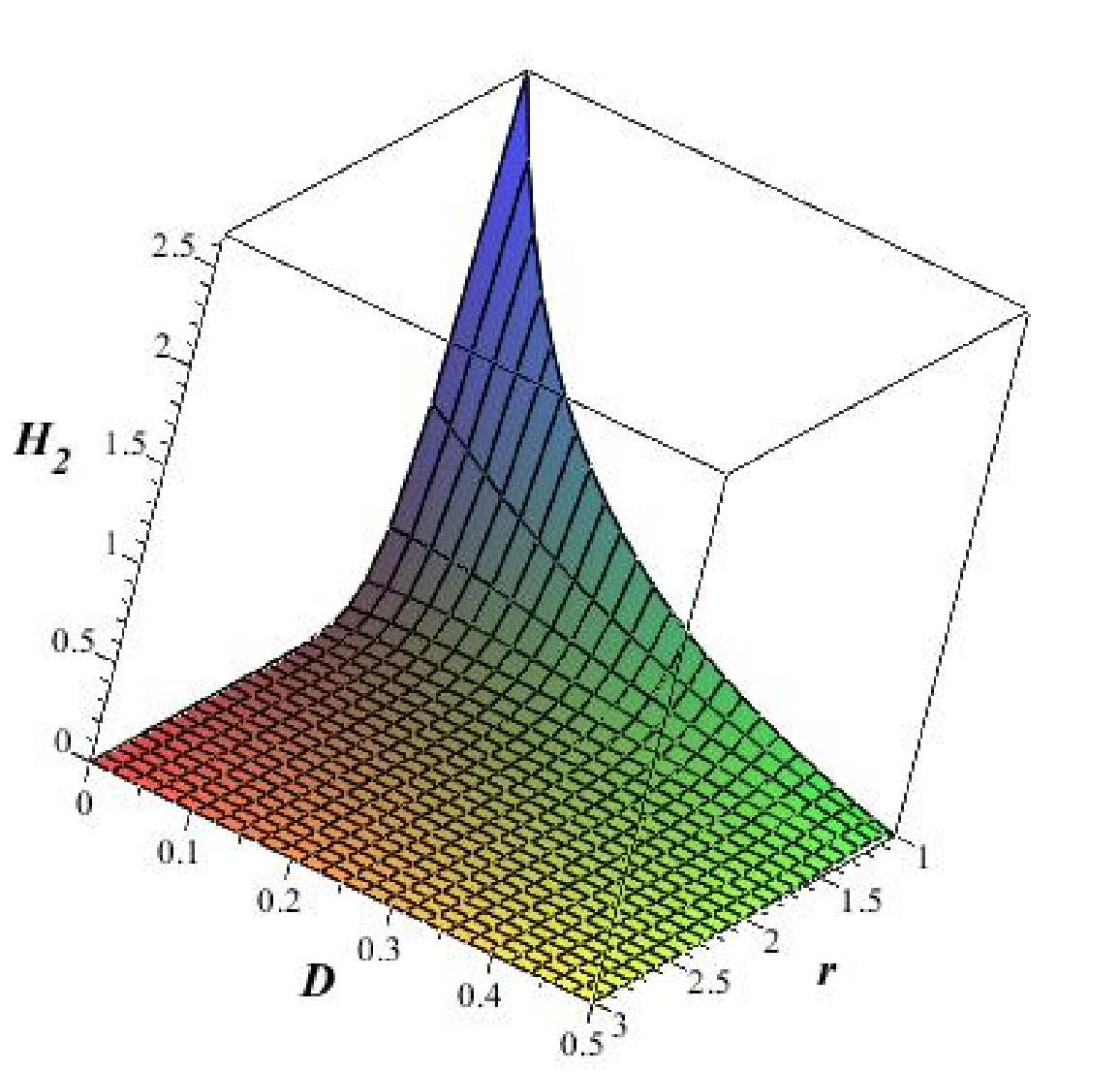}}\\
\subfloat[$H_3>0$]{\includegraphics[width = 2.5in]{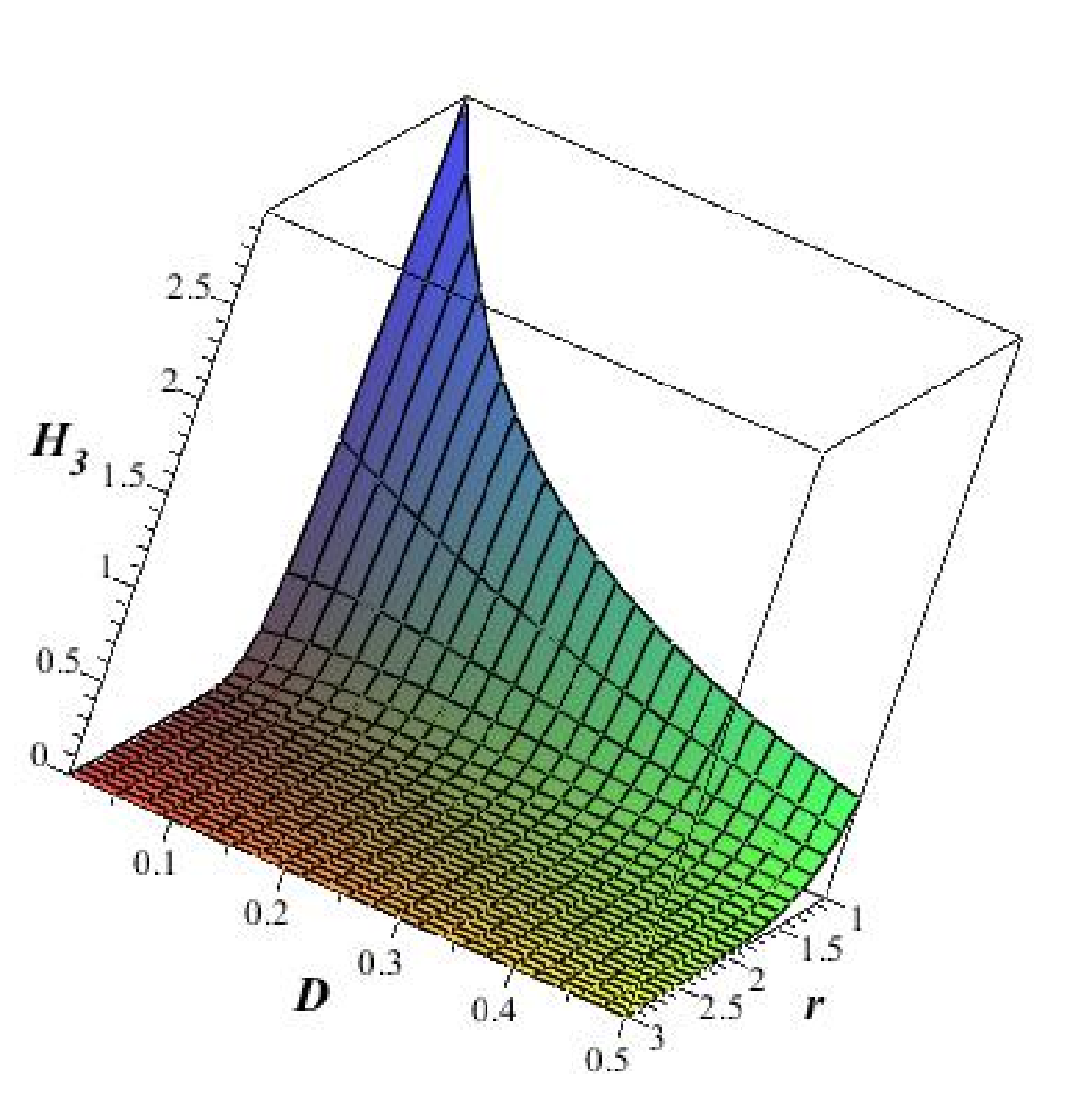}}\\
\subfloat[$H_4>0$]{\includegraphics[width = 2.5in]{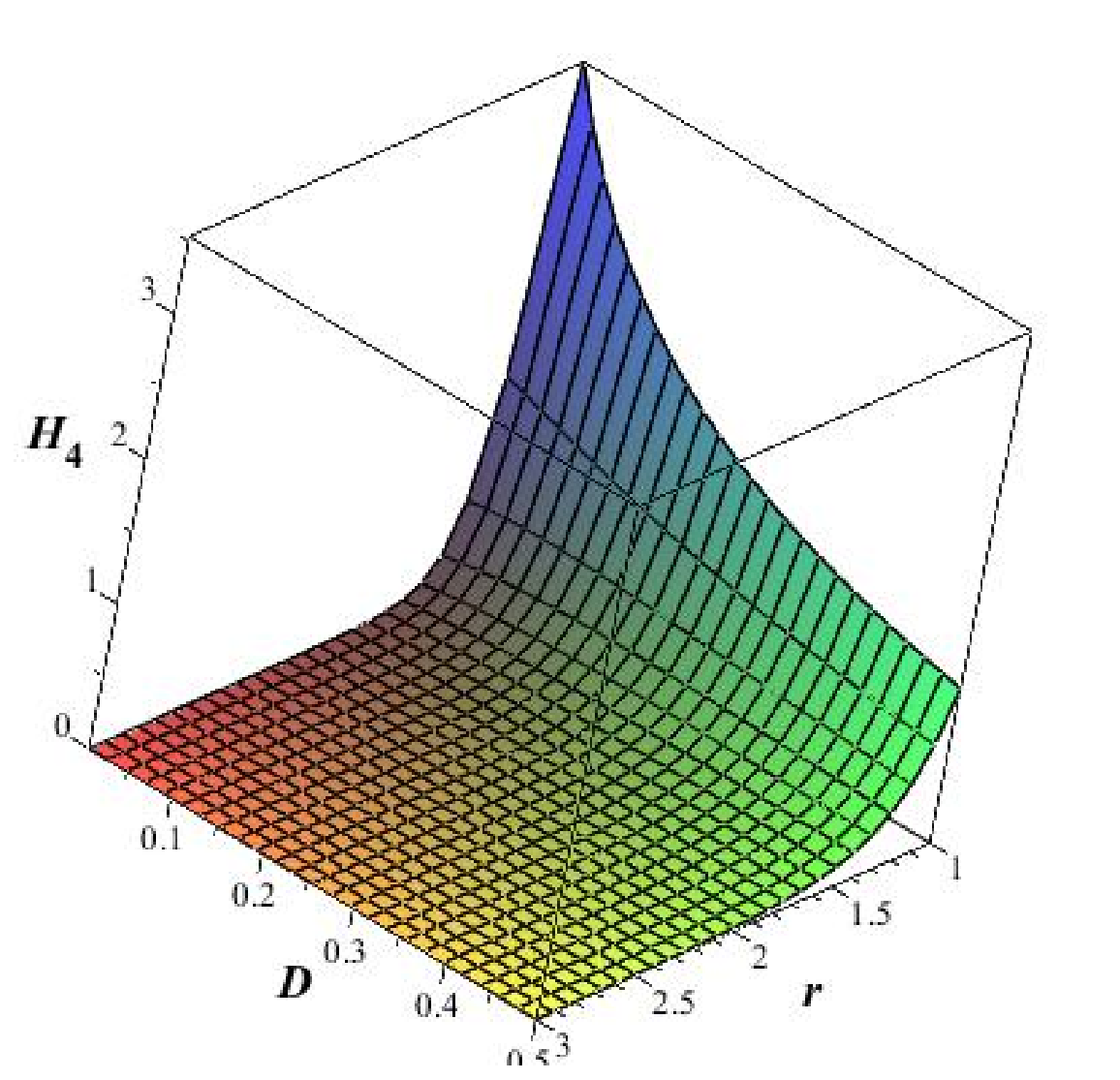}}
\caption{The graph depicts the correlation between $H_2(r, D)$ (a), $H_3(r, D)$ (b), and $H_4(r, D)$ (d) against  $r$ and $D$ for $\alpha=\beta=-1/2$. It is clear that all of these functions  are positive in the range $0<D<1/2$.}
\label{fig13}
\end{figure}

The energy density for this shape function is plotted as a function of $r$ and $D$ for two different ranges of $D$ in Fig. (\ref{fig9}). This figure indicates that positive energy density is accessible in the range $0<D$. In the subsequent phase of our analysis, we have illustrated $H(r,D)$ as a function of $D$ and the radial coordinate in Fig.(\ref{fig10}), revealing that $H>0$ can be attained within the interval $0<D$. However, the lateral NEC diagram (Fig.  (\ref{fig11})) indicates that the condition $H_1>0$ is satisfied only within a limited range of positive values for $D$. Figure (\ref{fig12}) verify that $H_1>0$ is valid in the range $0<D<1/2$. We have plotted $H_2(r,D)$, $H_3(r,D)$, and $H_4(r,D)$ in the range $0<D<1/2$ in Fig. (\ref{fig13}). Now, we can conclude that the shape function (\ref{37}) satisfies all ECs in the range $0<D<1/2$.

\subsection{ Solution with known shape function }\label{subsec6}
A class of asymptotically flat wormhole solutions that satisfy both radial and lateral NEC is discussed in reference \cite{SR1}. It has been demonstrated that the function $f(R,T)=R+2\lambda T$ yields solutions that satisfy the ECs. The inclusion of the term $2\lambda T$ in the $f(R,T)$ function results in a modification of the Einstein gravitational constant to $1+2\lambda $ \cite{SR1}. The presence of a negative value for this effective gravitational constant may provide a resolution to the problem of exotic matter. In this subsection, we examine some of the solutions outlined in \cite{SR1} within the framework of $f(T,\mathcal{T})$ gravity. As an example, we consider
\begin{equation}\label{209b}
b(r)=(C +D)^{m}(C r+D)^{m}.
\end{equation}

This shape function is discussed in \cite{SR1} within the context of the $f(R,T)=R+2\lambda T$ theory, specifically for a variable EoS where $\omega(r)=\omega_\infty+D/r$ is taken into account. In that scenario,
\begin{equation}\label{210b}
C=\omega_\infty,\quad m=-1/\omega_\infty.
\end{equation}
It is easy to show that the shape function (\ref{209b}) in the background of $f(T,\mathcal{T})$ gives
\begin{equation}\label{211b}
\omega(r)=\frac{p(r)}{\rho (r)}=\omega_\infty+g(r),
\end{equation}
where
\begin{equation}\label{212b}
\omega_\infty=\frac{3m\beta+4-5\beta}{m\beta-4m+\beta},
\end{equation}
and
\begin{equation}\label{212c}
g(r)=\frac{8Dm(\beta^2-3\beta+2)}{(Cr(m(\beta-4)+\beta)+\beta D)(4m-m\beta-\beta)}.
\end{equation}
One can use (\ref{212b}) to find
\begin{equation}\label{213b}
m=\frac{4-5\beta-\omega_\infty \beta}{\omega_\infty\beta-3\beta-4\omega_\infty}.
\end{equation}

\begin{figure}
\subfloat[$\rho>0$]{\includegraphics[width = 2in]{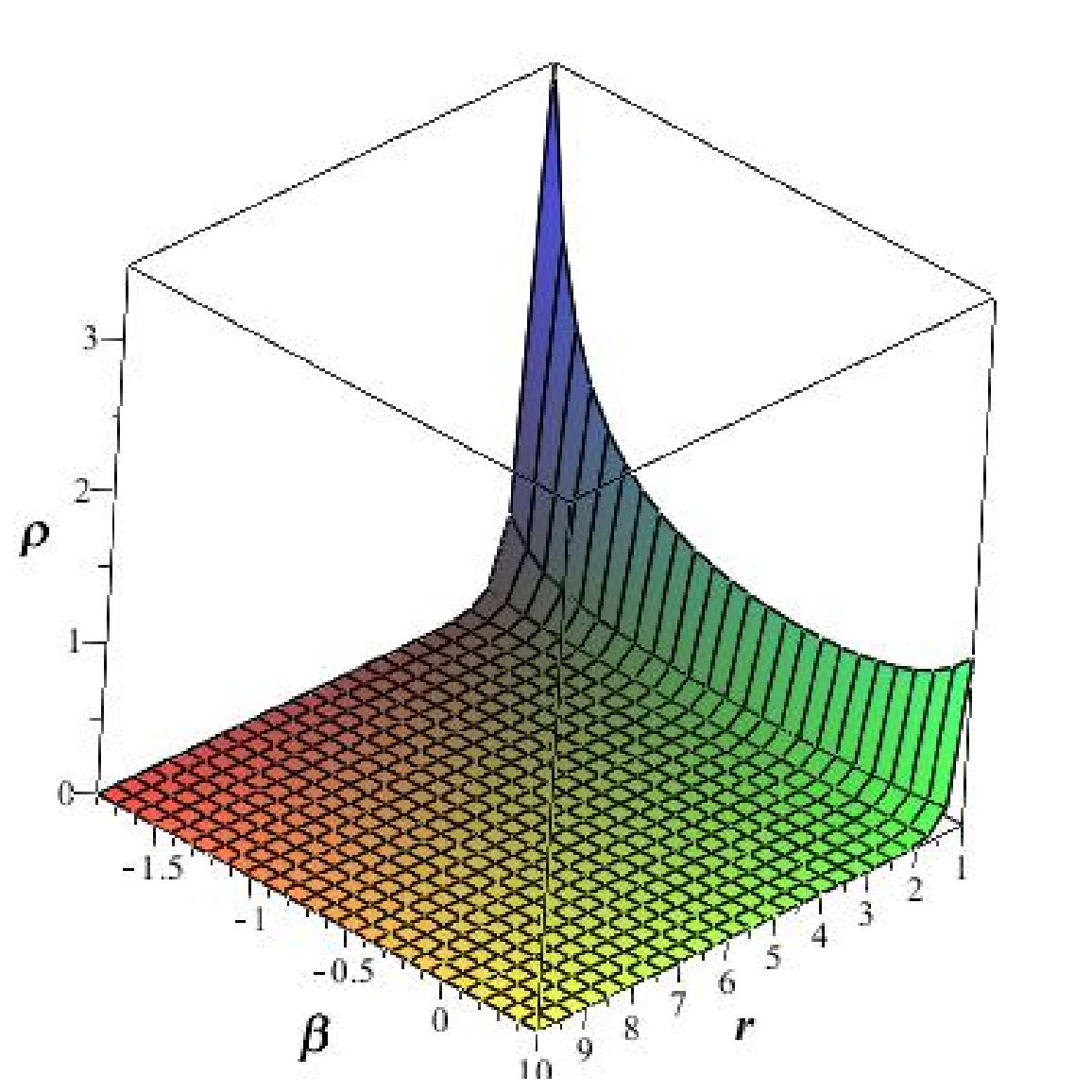}}\\
\subfloat[$H>0$]{\includegraphics[width = 2in]{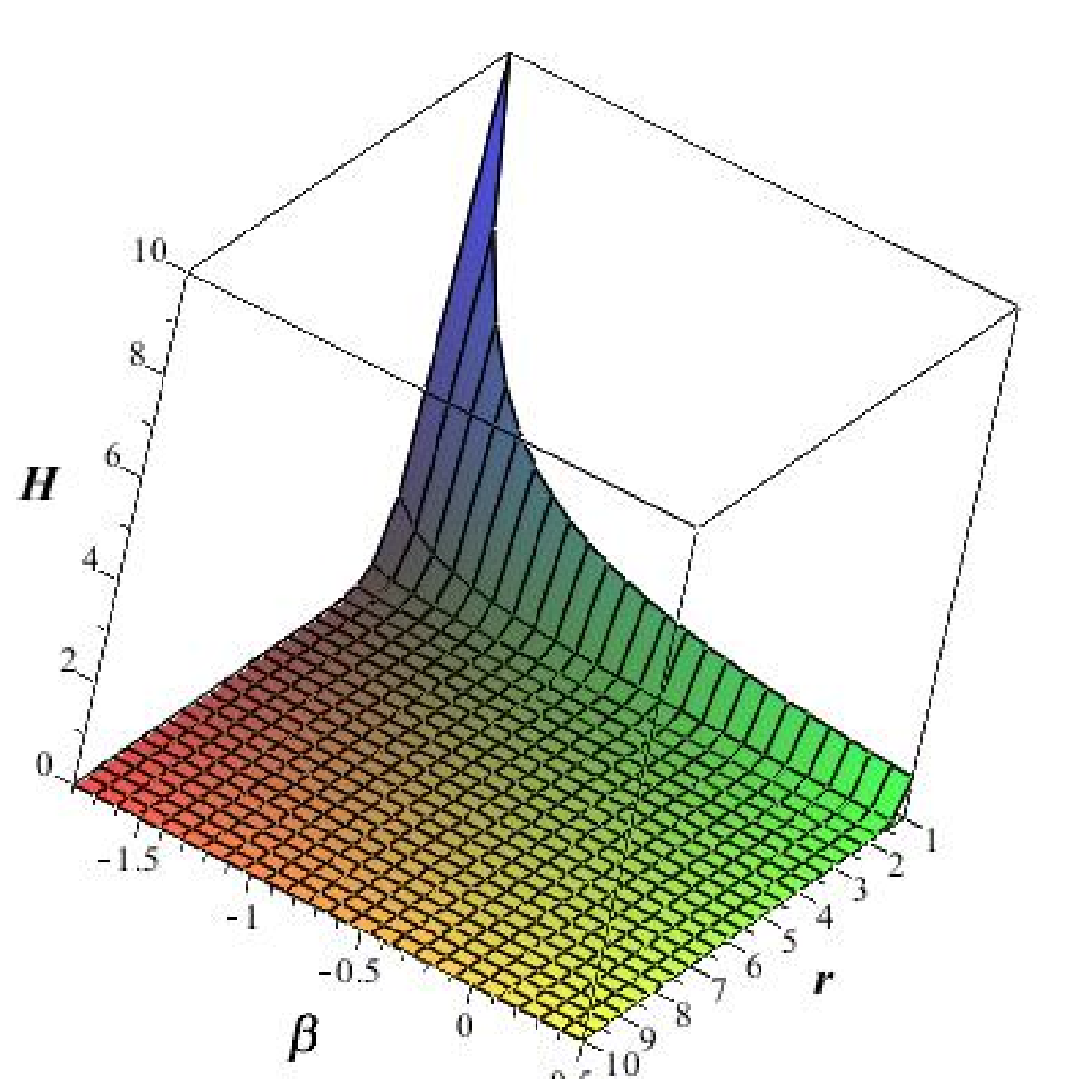}}\\
\subfloat[$H_1>0$]{\includegraphics[width = 2in]{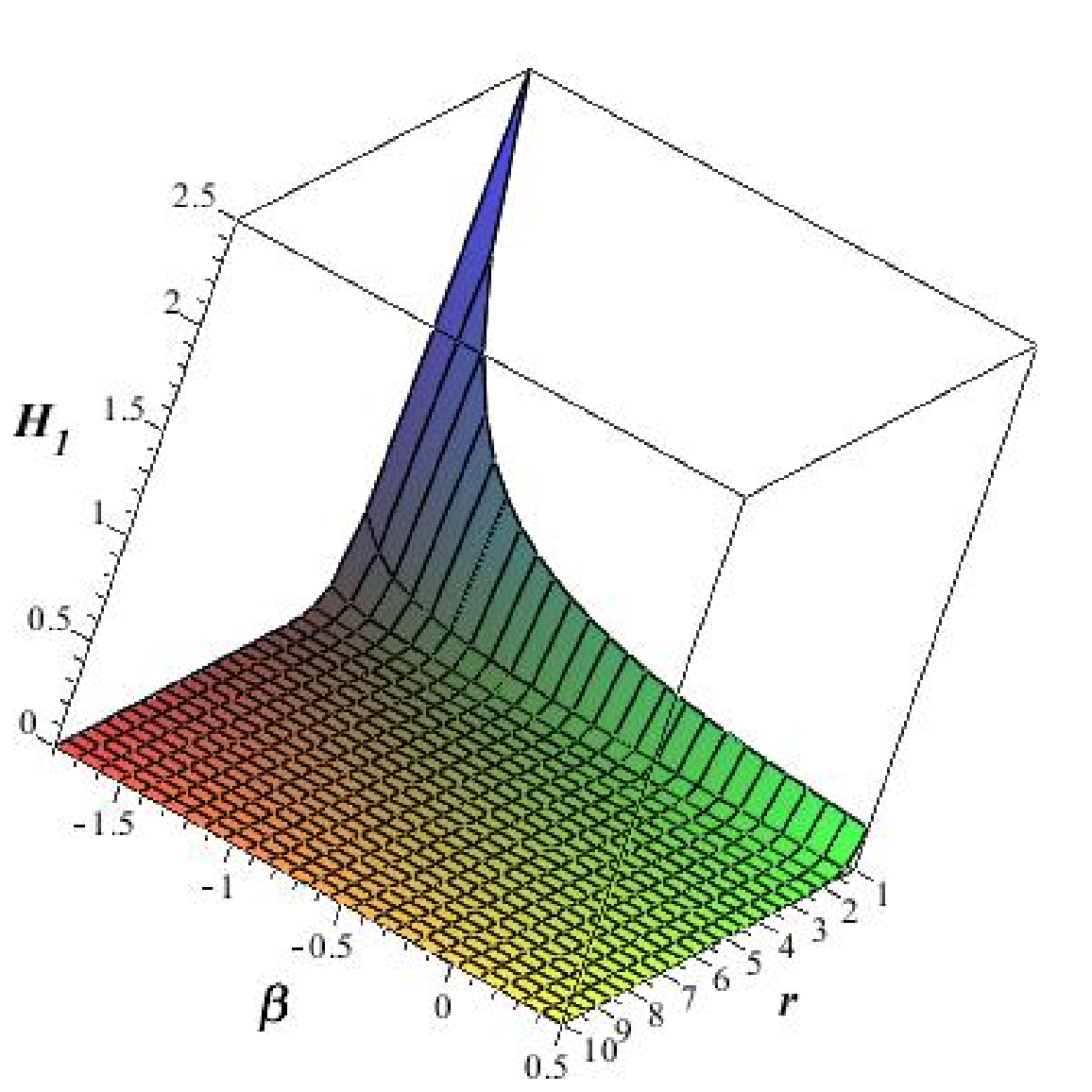}}\\
\subfloat[$H_3>0$]{\includegraphics[width = 2in]{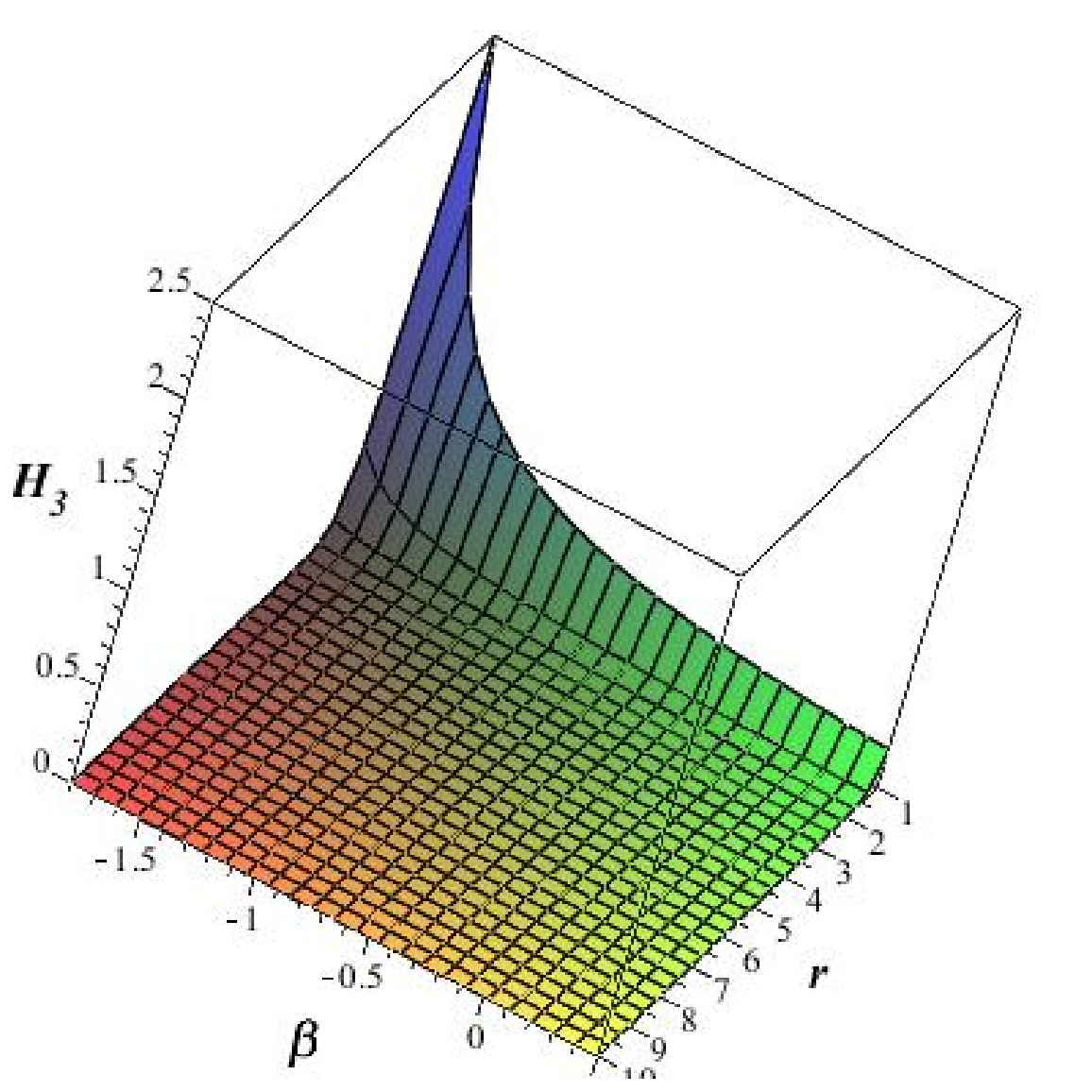}}
\caption{The graph depicts the correlation between $\rho(r, \beta)$ (a), $H(r, \beta)$ (b), $H_1(r, \beta)$ (c), and  $H_4(r, \beta)$ (d) against  $r$ and $D$. It is clear that all of these functions  are positive in the range $-2<\beta<0.5$.}
\label{fig14}
\end{figure}

The shape function (\ref{209b}) simultaneously contravenes the radial and lateral NEC within the framework of GR; so, it adheres to the NEC in the context of $f(T,\mathcal{T})$ theory. The examination of other ECs within the $f(T,\mathcal{T})$ theory is contingent upon various parameters ($C, D, \alpha, \beta$, and  $m$). To simplify our analysis, we will focus on the ECs while treating some of these parameters as constant. As an example, we set $C=D=1, \alpha=-1/2$ and $m=-6$ which results in
\begin{equation}\label{214b}
b(r)=\frac{64}{(1+r)^{6}}.
\end{equation}
This shape function gives
\begin{equation}\label{215b}
\rho(r,\beta)=\frac{8(5r\beta-24r-\beta)}{(\beta+2)(\beta-1)(r+1)^{7}r^3}.
\end{equation}

\begin{figure}
\subfloat[$H_2$]{\includegraphics[width = 2.4in]{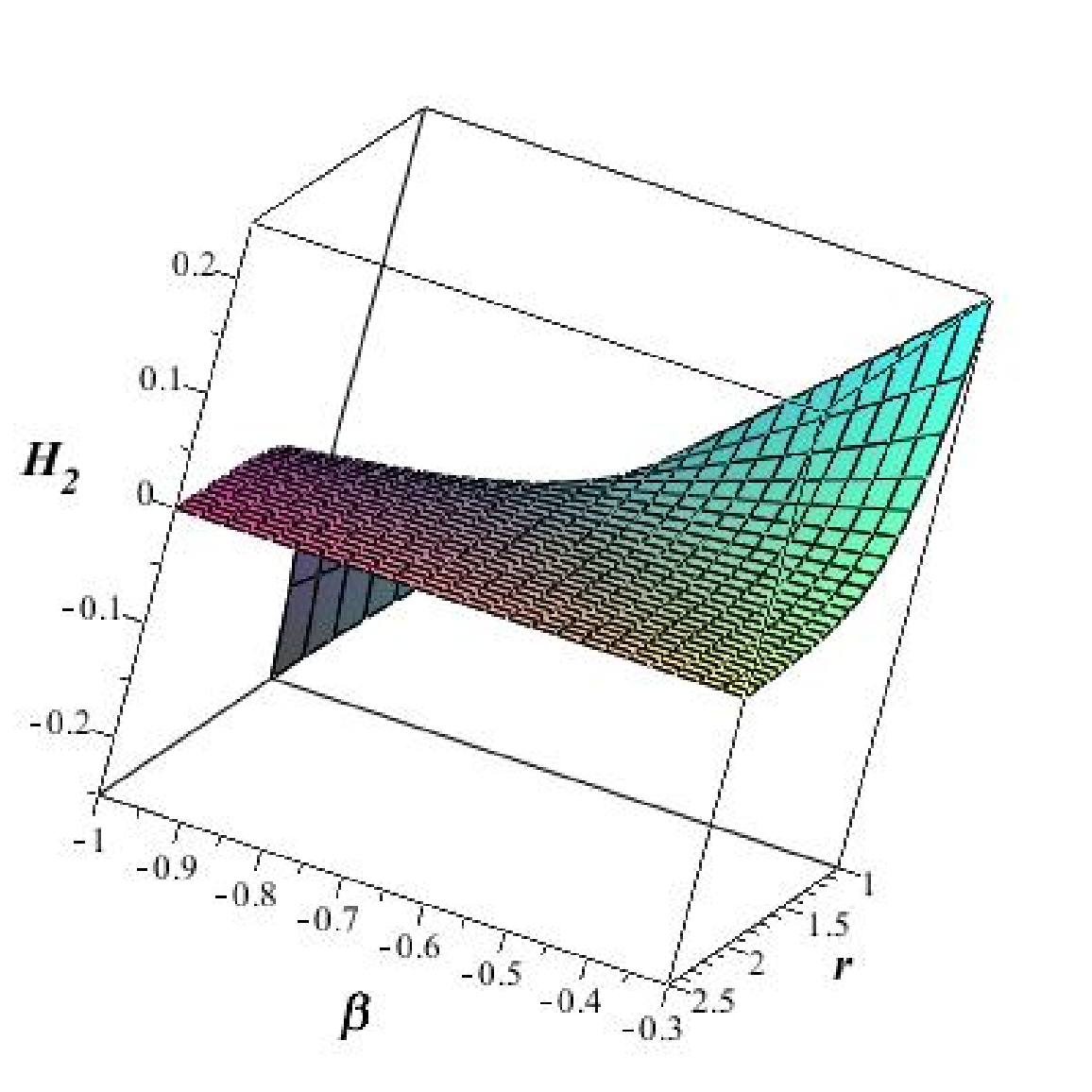}}\\
\subfloat[$H_4$]{\includegraphics[width = 2.4in]{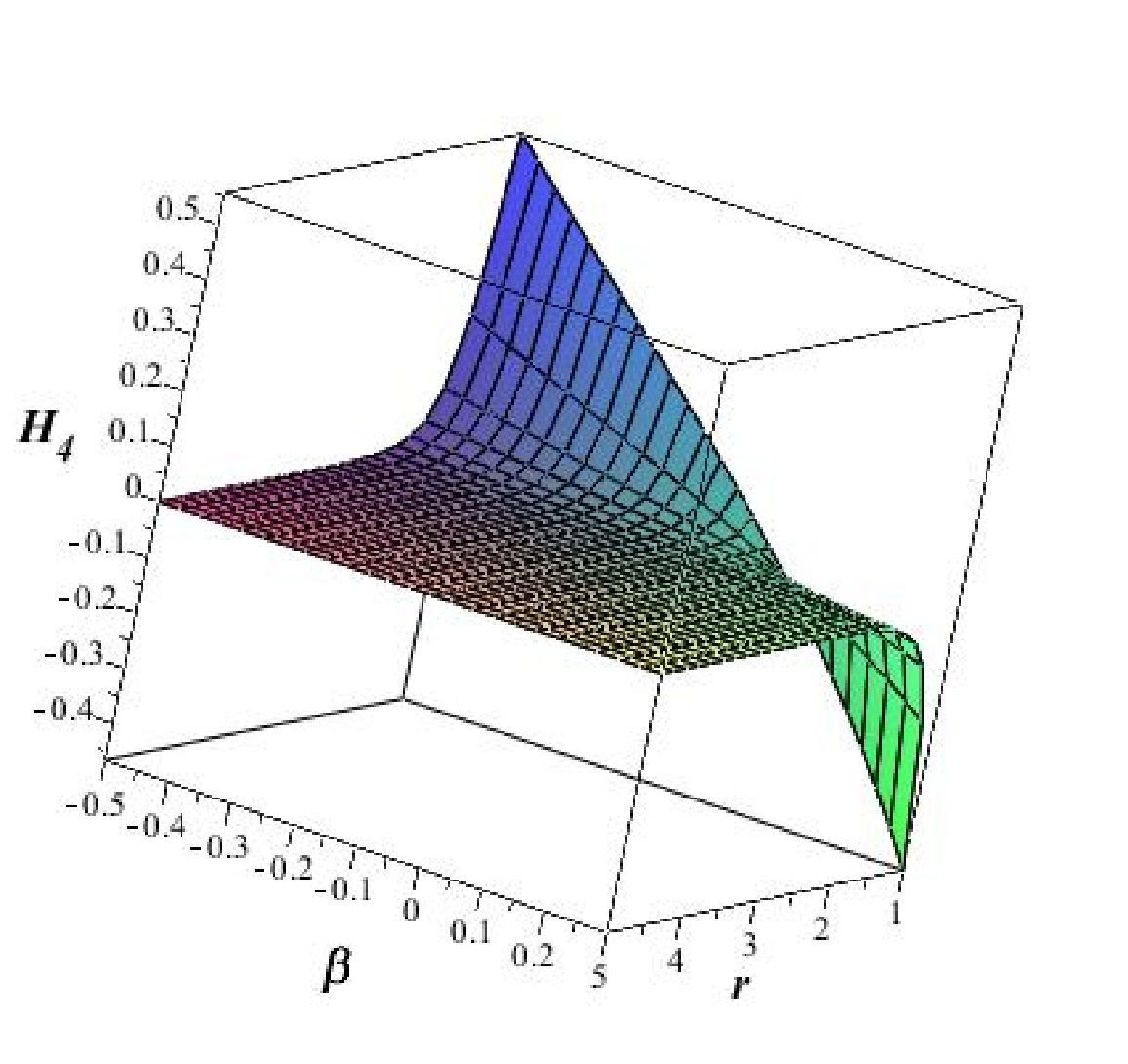}}
\caption{The graph illustrates $H_2(r, \beta)$ (a) and $H_4(r, \beta)$ (b) in relation to $r$ and $\beta$for $\alpha=-1/2$. It is evident that both of these functions exhibit a change in sign within the specified range.}
\label{fig15}
\end{figure}

\begin{table*}[]
\begin{tabular}{|l|l|l|l|}
\hline
$\quad \quad \beta$ & $H_2>0$ & $H_4>0$  & \quad Satisfied ECs  \\ \hline
 $0.1<\beta<0.5$  &  \quad\checkmark  & \quad $\times$ &  $NEC$,$WEC$,$DEC$\\ \hline
  $-0.5<\beta<0$ &\quad \checkmark  & \quad\checkmark &  $NEC$,$WEC$, $DEC$, $SEC$  \\ \hline
$-2<\beta<-1$ & \quad $\times$  & \quad\checkmark &$NEC$,$WEC$, $SEC$ \\ \hline
\end{tabular}
\caption{The results for three individual range of $\beta$ for $\alpha=-1/2$ while $b(r)=\frac{64}{(r+1)^{6}}$  is considered.  }\label{fe5}
\end{table*}

We have plotted $\rho, H, H_1$ and $H_3$ as a function of $r$ and $\beta$ in Fig. (\ref{fig14}). This figure indicates that all of these functions are positive in the ranges $r>1$ and $-2<\beta<0.5$. On the other hand, The $H_2$ and $H_4$ as a function of $r$ and $\beta$ are depicted in Fig. (\ref{fig15}). This figure indicates that $H_2$ and $H_4$ change the sign for different range of $\beta$. Given that all functions, except for $H_2$ and $H_4$, are positive within the interval $-2<\beta<0.5$, we present a summary of the overall behavior of these functions in Table (\ref{fe5}), along with the status of the ECs for three individual regions.
We now examine wormhole solutions within the framework of the curvature-based theory represented by $ f(R,T) = R + 2\lambda T$ and the torsion-based theory denoted as $ f(T,\mathcal{T}) = \alpha T + \beta \mathcal{T}$. Our findings indicate that solutions that satisfy both radial and lateral NEC in the context of GR can also comply with the ECs in both $ f(R,T)$ and $ f(T,\mathcal{T})$. However, the adherence to these ECs is explicitly dependent on the parameters $\alpha$ and $\beta$. Additionally, it is noteworthy that the same metric yields distinct energy-momentum tensors and, as a result, different EoS in these two theoretical frameworks.

\section{Concluding remarks}

In recent decades, the investigation of wormhole geometry within scientific research has generated significant excitement among scholars. In the context of GR, the material composition of wormholes is required to be exotic, necessitating a violation of certain ECs and the presence of negative mass. Consequently, modified theories of gravity may provide solutions to the challenges associated with exotic objects. In recent years, considerable advancements have been achieved in the modified theories of gravity, as researchers explore a range of extensions to GR.

A set of broader geometrical frameworks beyond Riemannian geometry, which could be applicable at the scale of the solar system, may offer insights into the behavior of matter on a cosmic scale. There has been an increasing interest in teleparallel gravity, an alternative form of modified gravity that employs torsion rather than curvature. Motivated by the teleparallel formulation of GR, we endeavored to explore the extension of $f(T)$ gravity through the interaction between the torsion scalar $T$ and the trace of the energy-momentum tensor $\mathcal{T}$. A notable extension is $f(T,\mathcal{T})$  gravity, which alters the conventional Einstein-Hilbert action by substituting the Ricci scalar with a function that incorporates both the torsion scalar and the trace of the energy-momentum tensor. The straightforward nature of the resulting field equations from the torsional perspective, which are of second order, makes it considerably more logical to explore the interaction between gravitation and matter via torsion instead of curvature, as the latter would yield fourth-order field equations. In the modified theories of gravity, the exploration of wormhole solutions emerges as an increasingly intriguing subject, owing to the incorporation of an effective energy-momentum tensor that leads to a violation of the NEC, independent of the existence of any distinct exotic matter.

In this work, we constructed new classes of asymptotically flat wormhole solutions in the context of $f(T,\mathcal{T})$ gravity which respects the ECs. Our findings indicate that non-exotic wormhole solutions can exist in this context. The function $f(T,\mathcal{T})$ is obtained from the tetrad field. Reference \cite{Tetrad} introduced two different categories of tetrads, namely diagonal and off-diagonal tetrads. In this research, we utilize the diagonal tetrad formalism. In this instance, it is permissible to develop the spherical solutions with confidence.

Initially, we derived the fundamental field equations, which, in contrast to GR, incorporate modifications that are contingent upon the two coupling parameters of the theory. Then we have shown that the violation of NEC depends on the sign of $Y(\alpha,\beta)$. It has been shown that solutions that violate both radial and lateral NEC within the framework of GR adhere to the NEC in the ranges (\ref{4cc}). The $f(T,\mathcal{T})$ modification offers a distinct perspective on gravity that markedly contrasts with other theories grounded in torsion or curvature. This theory of $f(T,\mathcal{T})$ sets itself apart from both $f(T)$ gravity and curvature-centric $f(R,T)$ gravity. It has been demonstrated that NEC is contingent upon the sign of $\frac{\alpha}{\beta+2}$. In other terms, geometric configurations that infringe upon both radial and lateral NEC within the framework of GR may still satisfy NEC in the context of $f(T,\mathcal{T})$, provided that the sign of $\frac{\alpha}{\beta+2}$ be negative. The violation of the WEC is contingent upon the sign of $\rho$. It has been noted that the energy density exhibits distinct forms for the same shape function within GR and the  $f(T,\mathcal{T})$ theory. Therefore, solutions that contravene the radial and lateral NEC within the framework of GR may violate the WEC in the context of $f(T,\mathcal{T})$ gravity, even if the sign of $\frac{\alpha}{\beta+2}$ be negative.

Considering a linear EoS results in a power-law shape function ($b(r)=r^m$). This finding is consistent for wormholes within the framework of GR; however, it is important to highlight that in GR, the parameter $m$ is proportional to $1/\omega$, whereas in the context of $f(T,\mathcal{T})$, it is influenced by $\omega$, and $\beta$ as indicated in Eq. (\ref{27}). It was demonstrated that $m\leq-1$ provides solutions that satisfy NEC, WEC, and SEC. We have also demonstrated that the DEC condition holds for certain values of $m$ and $\beta$, leading to the conclusion that $0 \leq \omega \leq 1$. It has been demonstrated that, within the framework of linear EoS, asymptotically isotropic wormhole solutions, as well as those that are approximately isotropic, are not attainable in the context of $f(T,\mathcal{T})$. A significant number of the shape functions presented in the literature do not simultaneously meet the radial and lateral NEC conditions \cite{SR1}. Consequently, we have employed a variable EoS method to develop non-exotic wormhole solutions in the background of $f(T,\mathcal{T})$. A diverse array of intriguing wormhole solutions was discovered, contingent upon the model parameters. A set of wormhole solutions that concurrently meet the radial and lateral NEC is introduced in \cite{SR1}. We have demonstrated that these solutions may adhere to the ECs in the context of $f(T,\mathcal{T})$ while utilizing a distinct energy-momentum tensor.

To deem a modified theory as a plausible candidate for elucidating natural phenomena, comprehensive investigations are imperative. In particular, it is vital to perform an in-depth comparison with cosmological observations, such as Baryon Acoustic Oscillations (BAO) and the Cosmic Microwave Background (CMB), along with the constraints imposed by Big Bang Nucleosynthesis (BBN), which could restrict the acceptable models and parameter ranges. Furthermore, after deriving spherically symmetric solutions, it would be advantageous to assess $f(T,\mathcal{T})$ gravity concerning data obtained from the Solar System.

\end{document}